\documentclass[final,5p,times,twocolumn]{elsarticle}

\usepackage{amssymb}
\usepackage{lipsum}
 \usepackage{amsthm}

\usepackage[T1]{fontenc}
\usepackage[utf8]{inputenc}

\usepackage{amsmath}
\usepackage{amssymb}

\usepackage{graphicx}
\usepackage[caption=false]{subfig}
\graphicspath{{./}{figures/}}

\usepackage{booktabs}
\usepackage{array}
\usepackage{multirow}
\usepackage{makecell}
\usepackage{tabularx}
\usepackage{longtable}
\usepackage{pdflscape}
\usepackage{placeins}

\usepackage{xurl}
\usepackage[hidelinks]{hyperref}
\usepackage[switch,mathlines]{lineno}
\biboptions{numbers,sort&compress}
\newcommand{\arcsec}{\ensuremath{^{\prime\prime}}}

\newcommand{\tablefoot}[1]{%
\par\vspace{1mm}
\begin{minipage}{0.98\linewidth}
\footnotesize
\textit{Notes.} #1
\end{minipage}
}

\journal{}
\makeatletter
\def\ps@pprintTitle{%
  \let\@oddhead\@empty
  \let\@evenhead\@empty
  \let\@oddfoot\@empty
  \let\@evenfoot\@empty
}
\makeatother

\begin{document}

\begin{frontmatter}

\title{Optical constraints on intranight variability in TeV HBLs from
a two-year Asiago Schmidt monitoring campaign\tnoteref{obsnote}}

\tnotetext[obsnote]{Based on observations collected at the Asiago Schmidt
67/92 telescope (Asiago Mount Ekar, Italy), operated by
INAF -- Osservatorio Astronomico di Padova.}

\author[aff1]{Domenico Impiombato\corref{cor1}}
\ead{domenico.impiombato@inaf.it}

\author[aff2]{Giacomo Bonnoli}

\author[aff3]{Elisa Prandini}

\author[aff1]{Luca Zampieri}

\author[aff4]{Irene Albanese}

\cortext[cor1]{Corresponding author.}

\affiliation[aff1]{
organization={INAF, Osservatorio Astronomico di Padova},
addressline={Vicolo dell'Osservatorio 5},
postcode={I-35122},
city={Padova},
country={Italy}
}

\affiliation[aff2]{
organization={INAF -- Osservatorio Astronomico di Brera},
addressline={Via Brera 28},
postcode={I-20121},
city={Milano},
country={Italy}
}

\affiliation[aff3]{
organization={Dipartimento di Fisica e Astronomia "Galileo Galilei"  Universit\`a degli Studi di Padova},
addressline={Via Marzolo 8},
postcode={I-35131},
city={Padova},
country={Italy}
}

\affiliation[aff4]{
organization={Centro di Ateneo di Studi e Attivit\`a Spaziali ``Giuseppe Colombo''},
addressline={Via Venezia 1},
postcode={I-35131},
city={Padova},
country={Italy}
}

\begin{abstract}
We present a two-year homogeneous optical monitoring campaign of
three TeV-emitting high-synchrotron-peaked BL Lac objects,
1ES~1959+650, H~1426+428, and PG~1553+113, carried out with
the Asiago Schmidt 67/92~cm telescope. TeV blazars are key targets in high-energy astrophysics, relevant
for relativistic jet physics, particle acceleration, multiwavelength
variability studies, and, in selected cases, multimessenger programmes. 
In this context, dense optical monitoring
provides a low-energy synchrotron tracer of jet activity over
temporal baselines that are difficult to cover continuously with
high-energy facilities.

The dataset consists of more than 17,000 $R$- and $I$-band
photometric measurements obtained over 957 source--filter
observing sessions between 2024 February 1 and 2026 January 31,
for a total monitoring time of about 420~h. We applied a uniform
differential-photometry and classical F-test framework based on
two independent comparison-star pairs.

Robust IDV is detected only in H~1426+428, with 15 out of
43 analysed sessions classified as variable in the $R$ band and
5 out of 40 in the $I$ band. No significant IDV is detected in
1ES~1959+650 or PG~1553+113 within the adopted statistical
framework. For H~1426+428, the corresponding duration-weighted
duty cycles, computed using only variable sessions with a measured
local characteristic timescale, are
$DC_{{\rm dur},R}=28.0\%$ and $DC_{{\rm dur},I}=7.0\%$.
The inferred local extrapolated doubling/halving timescales are
of the order of 5--15~min in the observer frame.

These results indicate that rapid optical synchrotron variability
in TeV HBLs is intermittent and state-dependent. They also show
that robotic small- and medium-aperture optical facilities can
provide valuable long-term support for high-energy monitoring
programmes by supplying dense, uniform, and repeatable optical
coverage.
\end{abstract}


\end{frontmatter}

\section{Introduction}
\label{introduction}

Blazars are an extreme subclass of active galactic nuclei
(AGN), characterized by relativistic jets closely aligned with
the observer's line of sight \citep{Urry1995}. Relativistic
beaming amplifies their non-thermal emission and makes them
highly variable sources over a broad range of timescales and
frequencies, from radio wavelengths up to TeV gamma rays
\citep{Fossati1998,Donato2001,Ghisellini2010}.

Their broadband spectral energy distribution is commonly
interpreted in terms of two broad components. In leptonic
scenarios, synchrotron radiation from relativistic electrons
dominates the low-energy component, extending from radio to
UV/X-ray frequencies, while inverse-Compton scattering accounts
for the high-energy component extending into the gamma-ray
regime and, in some sources, up to TeV energies
\citep{Fossati1998,Ghisellini2010}.

TeV blazars are key targets in high-energy astrophysics. They
provide laboratories for studying relativistic jet physics,
particle acceleration, radiative cooling, and the connection
between compact emitting regions and large-scale jet activity.
They are also relevant in broader contexts, since very-high-energy
gamma-ray observations of blazars can be used to investigate
the propagation of gamma rays through the extragalactic
background light and to probe magnetized extragalactic
environments \citep{Costamante2002,Bonnoli2015}.

Variability is one of the most direct observational probes of
the compact emitting regions in blazar jets. Changes observed
on timescales from years down to hours provide constraints on
the size, structure, and physical conditions of the emitting
plasma, as well as on the particle acceleration and cooling
processes operating in the jet \citep{Wagner1995,Ulrich1997}.
For this reason, time-domain studies are central to the
interpretation of blazar emission across the electromagnetic
spectrum.

TeV-detected high-synchrotron-peaked BL Lac objects (HBLs)
are particularly relevant for this type of study. In these
sources, the synchrotron peak is located in the UV/X-ray
domain, while the optical band samples the lower-energy part
of the same synchrotron component associated with the
relativistic jet. Optical monitoring therefore provides a
low-energy tracer of jet activity that is complementary to
X-ray and gamma-ray observations
\citep{Padovani1995,Costamante2002,Bonnoli2015,Abdo2010sed}.
Although optical data alone do not provide a complete
multiwavelength description of the source state, they can
supply dense and long-term temporal coverage that is often
difficult to obtain continuously at high energies.

The study of TeV blazars is now intrinsically multiwavelength
and, in selected cases, also connected to multimessenger
programmes. In this framework, optical monitoring remains
important on all timescales: long-term observations establish
the baseline source state, while intranight observations test
the presence of compact and rapidly evolving synchrotron
components. Such information is useful for interpreting
high-energy activity and for planning coordinated follow-up
observations with X-ray and gamma-ray facilities.

Blazar variability spans a wide range of temporal regimes,
from long-term changes over months or years down to rapid
flux variations occurring within a single night. Intranight
variability (IDV; also referred to as intraday variability)
has been detected in several blazars and is commonly
interpreted as the result of compact and rapidly evolving
emission regions within relativistic jets, possibly associated
with shocks, turbulence, or localized particle acceleration
events \citep{Wagner1995,Ulrich1997}.

Recent optical monitoring studies have also reported IDV in
TeV HBLs. For example, \citet{Chang2024} detected intraday
variability in H~1426+428 during 7 out of 16 observing nights
using multicolour optical observations obtained between 2021
and 2023. Such detections indicate that rapid optical
synchrotron variability can occur in extreme HBLs, although
its incidence and persistence remain poorly constrained by
homogeneous long-term datasets.

The observational picture of optical IDV in BL Lac objects is
still heterogeneous. Some studies report statistically significant
intranight variability events
\citep{Racine1970,Carini1992,Meng2017,Fang2022}, whereas
other monitoring campaigns find little or no evidence for
secure optical IDV \citep{Bachev2012,Pandey2020}. This
diversity is particularly relevant for TeV-detected blazars,
for which the literature includes both clear optical IDV
detections and campaigns showing an absence of significant
variability on hourly timescales \citep{Pandey2020,Su2024}.

A major difficulty in optical IDV studies is that low-amplitude
variability can be affected by instrumental and observational
systematics. Seeing variations, photometric instabilities, image
motion on the detector, and the choice of comparison stars may
all generate spurious variability signatures if not carefully
controlled
\citep{Cellone2007,Bachev2012,deDiego2010}. For this reason,
homogeneous observing strategies, stable comparison-star
sequences, and conservative statistical criteria are essential
when estimating the incidence of secure IDV events.

Long-term optical monitoring with the same telescope and
instrumental configuration is therefore complementary to
coordinated multiwavelength campaigns. Strict simultaneity with
X-ray and gamma-ray facilities is crucial for studying individual
high-energy states, but such campaigns usually cover limited
time intervals. Uniform optical monitoring over many observing
seasons instead allows one to distinguish persistent variability
properties from rare or state-dependent rapid events, and can
provide useful context for the planning and interpretation of
high-energy follow-up observations.

Robotic small- and medium-aperture optical telescopes can play
a specific role in this context. For long-term monitoring and
duty-cycle studies, the availability of observing time, cadence,
repeatability, and homogeneous observing conditions can be as
important as aperture. When operated within coordinated
programmes or broader monitoring networks, such facilities can
provide the temporal coverage needed to complement larger
high-energy observatories.

In this work, we present a two-year homogeneous optical
monitoring campaign of three TeV HBLs carried out between
2024 and 2026 with the Asiago Schmidt 67/92~cm telescope.
The selected sources are established TeV emitters, and the
optical observations probe the synchrotron component associated
with their relativistic jets. The aim is to quantify the incidence
of secure optical IDV in these sources using a uniform
differential-photometry and F-test framework based on two
independent comparison-star pairs. Both detections and
non-detections are used to constrain the intermittency of rapid
optical synchrotron variability in TeV HBLs. We also derive
duration-weighted duty-cycle estimates for sessions showing
secure IDV and a measurable local characteristic variability
timescale.

The campaign also provides an observational test of the
effectiveness of the Asiago Schmidt telescope, in its robotic
configuration, for high-energy time-domain support studies.
The continuation of the programme will extend the temporal
baseline, include additional sources of interest, and benefit from
increasing automation of the reduction and variability-analysis
workflow, with the aim of integration into broader
multiwavelength collaborations.

\section{Observations}

\subsection{Target selection}

The targets included in this monitoring programme were selected
according to criteria aimed at optimising both the scientific
return and the observational feasibility of a systematic search
for optical intranight variability. The sample was restricted to
three TeV-detected BL Lac objects, with the goal of constructing
a small but physically homogeneous set of sources suitable for
long-term optical monitoring.

All selected targets belong to the HBL class, characterised by
weak or absent emission lines and broadband non-thermal emission
extending up to very-high-energy gamma rays. These sources are
particularly suitable for optical variability studies because their
synchrotron component extends into the UV/X-ray domain, providing
a direct connection between optical and high-energy jet activity.

The targets were also selected to have sufficient optical
brightness to allow high-precision differential photometry with
the Asiago Schmidt 67/92~cm telescope using relatively short
exposure times. This requirement ensures adequate signal-to-noise
ratios while minimising atmospheric and instrumental effects that
could mimic or obscure genuine variability signals.

Visibility constraints represented an additional selection
criterion. The sources were chosen to be optimally observable
from the Asiago Observatory over a significant fraction of the
year, allowing long nightly visibility windows and repeated
observations across multiple observing seasons. Such conditions
are particularly important for intranight variability studies,
which benefit from continuous temporal coverage over several
hours within the same night.

Finally, all selected targets are well-known TeV blazars that
have been extensively observed at high and very-high energies
by both ground-based Cherenkov telescopes and space-borne
gamma-ray missions \citep{Hinton2009,Abdo2010sed}. Optical
monitoring therefore provides valuable complementary
information for multiwavelength investigations of the physical
processes governing relativistic jets.

\subsection{The selected sources}

\textbf{1ES~1959+650} is a nearby high-energy-peaked BL Lac
object located at a redshift of $z=0.048$ \citep{Perlman1996}.
It is one of the brightest TeV blazars in the X-ray and
very-high-energy gamma-ray bands
\citep{Padovani1995,Massaro2011}. The source exhibits strong
multi-band variability on timescales ranging from minutes to
months, particularly at X-ray and TeV energies
\citep{Giebels2002,Krawczynski2004,Aliu2013}. Several orphan
TeV flares, not accompanied by simultaneous X-ray activity,
have also been reported, making 1ES~1959+650 an important
target for testing synchrotron self-Compton emission models
\citep{Krawczynski2004,Boettcher2005}. In contrast, optical
variability appears generally weaker and slower than that
observed at higher energies
\citep{Tagliaferri2003,Kapanadze2012,Kapanadze2016}.

\textbf{PG~1553+113} is a TeV-detected HBL characterised by a
synchrotron peak located in the UV--X-ray domain
\citep{Aharonian2006}. The source has been extensively monitored
at optical wavelengths over the past decades and exhibits
variability on timescales ranging from intraday to multi-year
behaviour \citep{Pandey2019,Agarwal2021,Dhiman2023}. Its
redshift, initially constrained through upper limits
\citep{Mazin2007,Prandini2010}, is currently estimated to be
$z \simeq 0.43$ \citep{Danforth2010,Johnson2019}.
PG~1553+113 is also known for a possible quasi-periodic
modulation with a period of about 2.2 years in the gamma-ray
band, as observed by \textit{Fermi}-LAT
\citep{Ackermann2015,Tavani2018}.

\textbf{H~1426+428} is an extreme high-energy-peaked BL Lac
object with a synchrotron peak located at frequencies of the
order of $10^{18}$~Hz \citep{Costamante2001}. The source is
relatively faint in the optical band but bright at X-ray and
TeV energies, and is located at a redshift of $z=0.129$
\citep{Remillard1989}. H~1426+428 was among the first BL Lac
objects detected at very-high-energy gamma rays
\citep{Horan2002,Aharonian2002,Djannati2002}, and subsequent
multiwavelength observations revealed pronounced variability,
particularly at high energies \citep{Abdo2009}. Recent optical
monitoring studies have also reported statistically significant
intranight variability in this source \citep{Chang2024},
making H~1426+428 a particularly suitable target for
systematic optical IDV studies.

\subsection{Instrumentation and observing strategy}

The observations were carried out using the Schmidt 67/92\,cm
telescope at the Asiago Observatory (INAF–Osservatorio
Astronomico di Padova), located on Mount Ekar at an altitude
of approximately 1370\,m above sea level. The telescope is a
classical Schmidt reflector equipped with a 67\,cm diameter
corrector plate and a 91\,cm spherical primary mirror, with
a focal length of about 2150\,mm (f/3.2). This optical
configuration combines a wide field of view with high
throughput, making the instrument particularly well suited for
photometric monitoring programmes and optical variability
studies on short and intermediate timescales.

Following a series of instrumental upgrades completed in recent
years, the telescope currently operates in fully robotic mode,
ensuring high observational stability and good repeatability of
the instrumental conditions, both essential for long-term
monitoring campaigns.

The telescope is equipped with a scientific CCD camera based
on an ON Semiconductor KAF-16803 detector\footnote{\url{https://www.onsemi.com/download/data-sheet/pdf/kaf-16803-d.pdf}},
a full-frame CCD with an active array of
4096~$\times$~4096 pixels and a pixel size of
9~$\times$~9\,$\mu$m. The sensitive area of the detector is
36.8~$\times$~36.8\,mm$^2$, corresponding at the Schmidt
focal plane to a pixel scale of approximately
0.87~arcsec\,pixel$^{-1}$ (unbinned) and to a total field of
view of about 59~$\times$~59~arcmin$^2$. This large field
allows the simultaneous inclusion of several comparison stars,
which is particularly advantageous for differential photometry.

The detector features a full-well capacity of approximately
$1\times10^5$ electrons per pixel and operates with a typical
gain of $\sim$1.6~e$^{-}$\,ADU$^{-1}$, providing a wide
linear dynamic range. The read-out noise is approximately
9~e$^{-}$ RMS, adequate for high-precision photometric
applications even with relatively short exposure times. The
dark current is rendered negligible at the typical operating
temperatures thanks to the multi-stage thermoelectric cooling
system of the camera.

The KAF-16803 sensor employs transparent gate technology and
microlenses, ensuring high sensitivity in the optical domain.
The maximum quantum efficiency reaches approximately 60\%
at 550\,nm, while the spectral response remains high in the
550--800\,nm wavelength range, making the detector particularly
well suited for photometric observations in the $R$ and $I$
bands.

The observations were performed primarily using the $R$ and
$I$ photometric filters, selected to maximise the signal-to-noise
ratio of the targets and to reduce the impact of atmospheric
effects compared to shorter-wavelength bands. The use of these
filters also allows a direct comparison with numerous previous
studies of optical blazar variability
\citep{Zhang2017,Dhiman2023,Izviekova2022}.

The typical natural seeing at the Mount Ekar site is of the
order of $\sim$1.0--1.5~arcsec, with values below 1.0~arcsec
not uncommon under favourable observing conditions. Under such
conditions, the combination of the Schmidt optical quality,
detector pixel scale, and instrumental stability enables
reliable differential photometry and adequate sampling of
stellar profiles on intranight timescales.

The observing strategy was optimised for intranight variability
studies by adopting sequences of approximately 10--90
consecutive exposures in the same filter. Individual exposure
times ranged from 30 to 120~s, providing temporal sampling
suitable for investigating rapid optical variability on
timescales from tens of minutes to hours while maintaining
uniformity among different observing nights.

\section{Data reduction and photometry}

\subsection{Image processing}

The behaviour of the three TeV blazars at different flux
levels was investigated through systematic observations
carried out with the Schmidt telescope over a period of
approximately two years, between February 2024 and
January 2026.

A total of 5712 images were acquired for 1ES~1959+650,
4422 for H~1426+428, and 7363 for PG~1553+113 in the
optical $R$ and $I$ bands. The complete observing log,
including the number of observing nights in each filter,
is reported in Table~1.

The CCD reduction and photometric analysis were performed
through a semi-automatic procedure combining visual inspection
with dedicated routines from the \emph{Starlink Software Collection}
(SSC) \citep{Currie2014,Berry2022,Bell2024}, developed within
the UK Starlink project\footnote{\url{http://starlink.eao.hawaii.edu}}.
The SSC provides a comprehensive suite of tools for astronomical
data reduction, image visualisation, and photometric analysis.

A preliminary inspection of the images was carried out manually
using the \emph{Graphical Astronomy and Image Analysis} (GAIA)
tool\footnote{\url{https://starlink.eao.hawaii.edu/docs/sc17.htx/sc17.html}}.
Only frames characterised by reliable photometry, stable sky
background, correct source centring, and the presence of suitable
calibration stars within the field of view were retained for the
subsequent analysis. Images affected by poor weather conditions
or evident artefacts were discarded.

To remove localised defects, such as small artefacts or limited
portions of satellite trails close to the target field, an
\emph{image patching} procedure was applied. This method allows
selected image regions to be replaced with an interpolated surface
derived from the surrounding area, including an artificial noise
component. The use of this procedure was restricted to marginal
cases and applied with particular caution. Images in which
satellite trails or debris overlapped directly with the blazar
or comparison stars were excluded from the analysis.

The reduction of the raw CCD frames was performed using the
\emph{CCDPACK} software package\footnote{\url{https://starlink.eao.hawaii.edu/docs/sun139.htx/sun139.html}},
which provides standard routines for bias subtraction, dark-current
removal, flat-fielding, cosmic-ray rejection, image registration,
and normalisation.

\subsection{Aperture photometry}

Aperture photometry was performed on each reduced CCD image
using the \emph{PHOTOM} software package
\footnote{\url{https://starlink.eao.hawaii.edu/docs/sun45.htx/sun45.html}}.
This package allows both interactive photometry, useful for
detailed inspection of the targets and their immediate
environment, and the use of externally provided coordinate
files, enabling a reproducible analysis of large datasets.

For each observing night, photometry was carried out using
several aperture radii. The choice of the photometric aperture
was based on a compromise among multiple requirements:
maximising the signal-to-noise ratio, minimising the scatter
of the photometric measurements, and reducing the contribution
from the host galaxy emission, which is particularly relevant
for \emph{host-dominated} blazars \citep{Nilsson2007}.
Table~\ref{tab:obs_log} summarises the observing log of the
campaign, including the number of observing nights, the total
number of frames, the overall monitoring time, and the aperture
radii adopted for each source.

\begin{table}
\centering
\caption{
Observing log of the optical photometric observations.
The number of nights refers to source--filter observing nights.
Aperture radii are given in arcseconds.
}
\label{tab:obs_log}
\begin{tabular}{l c c c c c}
\hline\hline
Blazar & Filt. & Nights & Frames & Tot. & Apert. \\
\hline
1ES~1959+650 & R & 176 & 2895 & 80~h & 4\arcsec\  \\
             & I & 169 & 2817 & 76~h & 4\arcsec\  \\
H~1426+428   & R & 126 & 2294 & 43~h & 3\arcsec\  \\
             & I & 124 & 2128 & 40~h & 3\arcsec\  \\
PG~1553+113  & R & 184 & 3824 & 93~h & 6\arcsec\  \\
             & I & 178 & 3539 & 86~h & 6\arcsec\  \\             
\hline
\end{tabular}
\end{table}

The seeing during the observations typically varied between
approximately 1\arcsec\ and 2\arcsec.
In all cases, the photometric aperture was chosen to include
more than 90\% of the stellar flux within the aperture,
thereby mitigating the impact of seeing variations on the
photometric measurements.
The adopted apertures were also checked by inspecting the
dependence of the differential magnitudes on the measured
full width at half maximum (FWHM); no systematic trends were
found for the comparison-star light curves used in the IDV
analysis.

The comparison stars were selected according to criteria of
angular proximity, adequate signal-to-noise ratio, and
photometric stability as assessed from the star--star
differential light curves.

For 1ES~1959+650, comparison stars numbered 2, 4, 6, and 7
were adopted, chosen for their demonstrated stability
\citep{Pace2013}. For H~1426+428, comparison stars A, B, C,
and D were used, following the photometric sequence of
\citet{Smith1991}. Finally, for PG~1553+113, comparison
stars numbered 1, 2, 3, and 4 were selected from ground-based
observations conducted by the \emph{Whole Earth Blazar Telescope}
(WEBT) collaboration \citep{Raiteri2015}, calibrated in the
Bessell filter system \citep{Bessell1998}.

For the variability analysis, the four comparison stars in each
field are denoted generically as $S_1$, $S_2$, $S_3$, and $S_4$.
The correspondence between these generic labels and the
identifiers used in the original photometric sequences is given
in Table~\ref{tab:comparison_stars}. The adopted independent
comparison-star pairs are therefore $(S_1,S_2)$ and $(S_3,S_4)$
for all sources.

\begin{table*}
\centering
\caption{
Reference magnitudes and identifiers of the comparison stars used
for differential photometry.
The labels $S_1$--$S_4$ indicate the generic notation adopted in
the variability analysis, while the column ``Star ID'' gives the
identifiers used in the original finding charts and photometric
sequences: \citet{Pace2013} for 1ES~1959+650,
\citet{Smith1991} for H~1426+428, and \citet{Raiteri2015}
for PG~1553+113.
The uncertainties are reported in parentheses.
}
\label{tab:comparison_stars}
\begin{tabular}{l c c c c}
\hline\hline
Blazar &
Label &
Star ID &
$R$ (mag) &
$I$ (mag) \\
\hline
1ES~1959+650 & $S_1$ & 2 & 12.534 (0.004) & 12.217 (0.004) \\
              & $S_2$ & 4 & 14.038 (0.004) & 13.619 (0.004) \\
              & $S_3$ & 6 & 14.758 (0.004) & 14.365 (0.004) \\
              & $S_4$ & 7 & 14.738 (0.004) & 14.315 (0.004) \\
\hline
H~1426+428   & $S_1$ & A & 13.23 (0.02)  & 12.43 (0.02)  \\
              & $S_2$ & B & 14.17 (0.02)  & 13.79 (0.02)  \\
              & $S_3$ & C & 13.00 (0.02)  & 12.62 (0.03)  \\
              & $S_4$ & D & 15.20 (0.04)  & 14.82 (0.05)  \\
\hline
PG~1553+113  & $S_1$ & 1 & 13.465 (0.032) & 13.080 (0.056) \\
              & $S_2$ & 2 & 13.582 (0.029) & 13.230 (0.055) \\
              & $S_3$ & 3 & 15.277 (0.026) & 14.896 (0.051) \\
              & $S_4$ & 4 & 15.317 (0.029) & 14.893 (0.056) \\
\hline
\end{tabular}
\end{table*}

We note that the average calibrated magnitudes obtained in this
work for comparison star~A in the H~1426+428 field,
corresponding to $S_1$ in Table~\ref{tab:comparison_stars},
are $R = 13.40 \pm 0.01$ and $I = 12.57 \pm 0.01$.
These values are systematically fainter by approximately
0.1--0.15 mag than the reference values reported in the
photometric sequence of \citet{Smith1991}.

The origin of this offset is not fully clear. It may be related
to differences in the absolute photometric calibration, filter
response, colour terms, or standard-star realisation between
the present dataset and the reference photometric sequence.
We therefore regard this discrepancy as worthy of further
investigation in future calibration work.

However, this offset does not affect the intranight variability
analysis presented here. The F-test is based on differential
light curves computed within each observing session, and the
comparison-star light curves show good internal stability over
the monitoring period, with no evidence for systematic intranight
variations or seeing-related trends.
For this reason, the internally consistent photometric
measurements derived in this work were adopted for the
subsequent differential analysis.

\section{Variability analysis}

\subsection{Statistical tests}

The presence of IDV was investigated
using the classical F-test, originally introduced for stellar
variability studies \citep{Howell1988} and later systematically
discussed in the context of AGN microvariability by
\citet{deDiego2010}. The F-test compares the variance of the
differential light curve (DLC) of the blazar with that of a
reference DLC obtained from comparison stars observed within
the same CCD frame, allowing an assessment of whether the
observed scatter exceeds that expected from photometric noise
alone.

Although analysis of variance (ANOVA) is often regarded as
one of the most powerful methods for detecting low-amplitude
variability \citep{deDiego2010}, its application requires a
sufficiently large and homogeneous temporal segmentation of
the intranight light curves.

In the present monitoring programme, observations were
obtained by alternating consecutive exposure blocks in the
$R$ and $I$ filters in order to probe IDV simultaneously in
both bands. This observing strategy introduces regular
temporal gaps within each single-filter light curve and often
produces only a limited number of temporal groups per night.

To ensure a uniform statistical treatment across the full
dataset and among the three monitored blazars, the classical
F-test was therefore adopted as the primary method for IDV
detection.

The $\chi^2$ test was not employed because it strongly depends
on the accuracy of the photometric error estimates, which are
often underestimated in real CCD data, potentially leading to
an increased rate of false positives
\citep{deDiego2010,Goyal2013}. Similarly, the so-called
C-test was not adopted because it lacks a well-defined
statistical distribution, making its commonly used
significance thresholds unreliable
\citep{deDiego2010,Villforth2010}.

\subsection{Variability classification}

To minimise the risk of spurious IDV detections caused by
photometric instabilities affecting individual comparison stars,
each blazar light curve was analysed using two independent
pairs of comparison stars. As described in
Table~\ref{tab:comparison_stars}, the four comparison stars
in each field are denoted generically as $S_1$, $S_2$,
$S_3$, and $S_4$. The adopted comparison-star pairs are
therefore $(S_1,S_2)$ and $(S_3,S_4)$ for all sources.

For each pair, two F-test statistics were computed from the
corresponding differential light curves, following the
approach commonly adopted in optical microvariability studies.
For the first comparison-star pair we computed

\begin{equation}
F_{1,12}=
\frac{\mathrm{var}(\mathrm{BL}-S_1)}
{\mathrm{var}(S_1-S_2)},
\qquad
F_{2,12}=
\frac{\mathrm{var}(\mathrm{BL}-S_2)}
{\mathrm{var}(S_1-S_2)},
\end{equation}

while for the second pair we computed

\begin{equation}
F_{1,34}=
\frac{\mathrm{var}(\mathrm{BL}-S_3)}
{\mathrm{var}(S_3-S_4)},
\qquad
F_{2,34}=
\frac{\mathrm{var}(\mathrm{BL}-S_4)}
{\mathrm{var}(S_3-S_4)}.
\end{equation}

Here, $\mathrm{BL}$ denotes the blazar light curve and
$\mathrm{var}$ indicates the sample variance of the corresponding
DLC. For each comparison-star pair, we then defined a pair-level
F statistic as the lower of the two values,

\begin{equation}
F_{12}=\min(F_{1,12},F_{2,12}),
\qquad
F_{34}=\min(F_{1,34},F_{2,34}).
\end{equation}

This choice requires the variance excess to be present regardless
of which comparison star in the pair is used to construct the
blazar--star DLC. It therefore provides an internal consistency
condition within each comparison-star pair and reduces the
probability of spurious detections caused by the instability of
a single comparison star.

The critical values $F_{\rm crit}$ were computed for
$\nu=N-1$ degrees of freedom in both the numerator and
denominator distributions, where $N$ is the number of
photometric points in the observing session.

The adopted significance thresholds,
$\alpha=0.01$ and $\alpha=0.001$, correspond to confidence
levels of 99\% and 99.9\%, respectively. The variability
classification was defined operationally as follows:

\begin{itemize}

\item a night is classified as ``variable'' (V) if both
$F_{12}$ and $F_{34}$ exceed the critical value corresponding
to $\alpha=0.001$;

\item a night is classified as ``probably variable'' (PV)
if one pair exceeds the $\alpha=0.001$ critical value and
the other pair exceeds the $\alpha=0.01$ critical value;

\item in all other cases, the night is classified as
``non-variable'' (NV).

\end{itemize}

This classification scheme was designed to favour reliability
over sensitivity, since a secure variability detection requires
agreement between two independent comparison-star pairs.

\subsection{Detection limits}

The statistical power of the classical F-test increases
with the number of photometric data points available within
a single observing night \citep{deDiego2010}. In this work,
only nights with a number of data points $N \geq 20$
per filter were considered suitable for the IDV analysis.

For variability amplitudes typical of optical BL Lac IDV
and for the adopted significance thresholds, analytical
estimates discussed by \citet{deDiego2010} indicate that
the statistical power of the F-test increases substantially
with increasing $N$, while maintaining strict control over
false positives.

According to this criterion, 43 and 40 nights satisfy
$N \geq 20$ in the $R$ and $I$ bands, respectively,
for H~1426+428; 49 and 48 nights for 1ES~1959+650;
and 70 and 66 nights for PG~1553+113.
These nights constitute the effective dataset used for
the IDV analysis and duty-cycle computation.

The corresponding average monitoring duration ranges from
approximately $\sim3.5$~h for H~1426+428 to
$\sim6$~h for 1ES~1959+650, with PG~1553+113
showing intermediate values of $\sim3.7$--$3.8$~h.

With regard to ANOVA, only a fraction of the observing
nights satisfies the minimum requirements for sufficiently
robust temporal segmentation in each filter.
In particular, the number of temporal groups and the
corresponding group sizes vary significantly from source
to source owing to the filter-switching observing strategy
adopted during the monitoring campaign.

A systematic application of ANOVA would therefore introduce
methodological inhomogeneities across the dataset, since the
statistical sensitivity and grouping structure would differ
substantially among the three monitored blazars.
For this reason, and in order to ensure uniformity and
comparability of the analysis, the classical F-test was
adopted as the primary statistical tool.

The dense temporal sampling achieved during the monitoring
campaign nevertheless ensures a reliable comparison between
the blazar light curves and those of the comparison stars,
even in the presence of regular gaps introduced by the
alternating $R$/$I$ observing sequence.

\subsection{Variability amplitude}

For nights classified as variable, we quantified the
intra-night variability amplitude using the standard
estimator introduced by \citet{Heidt1996}:

\begin{equation}
\mathrm{Amp}(\%) =
100\,\sqrt{(m_{\max}-m_{\min})^{2}-2\sigma^{2}},
\end{equation}

where $m_{\max}$ and $m_{\min}$ are the maximum and
minimum differential magnitudes measured during the night.

The term $\sigma^{2}$ represents the mean squared
photometric uncertainty associated with these two extrema
and is defined as

\begin{equation}
\sigma^{2} =
\frac{1}{2}
\left(
\delta m_{\max}^{2}
+
\delta m_{\min}^{2}
\right),
\end{equation}

where $\delta m_{\max}$ and $\delta m_{\min}$ are the
corresponding photometric errors.

The variability amplitude is reported only for nights
satisfying the most stringent variability criterion (V)
and is used solely as a descriptive quantity rather than
as a detection parameter. When the quantity under the
square root becomes negative, the variability amplitude
is not defined and is therefore set to zero.

\section{Results}

\subsection{Long-term optical behaviour}

\begin{figure*}
\centering

\includegraphics[width=\textwidth,height=0.28\textheight,keepaspectratio]{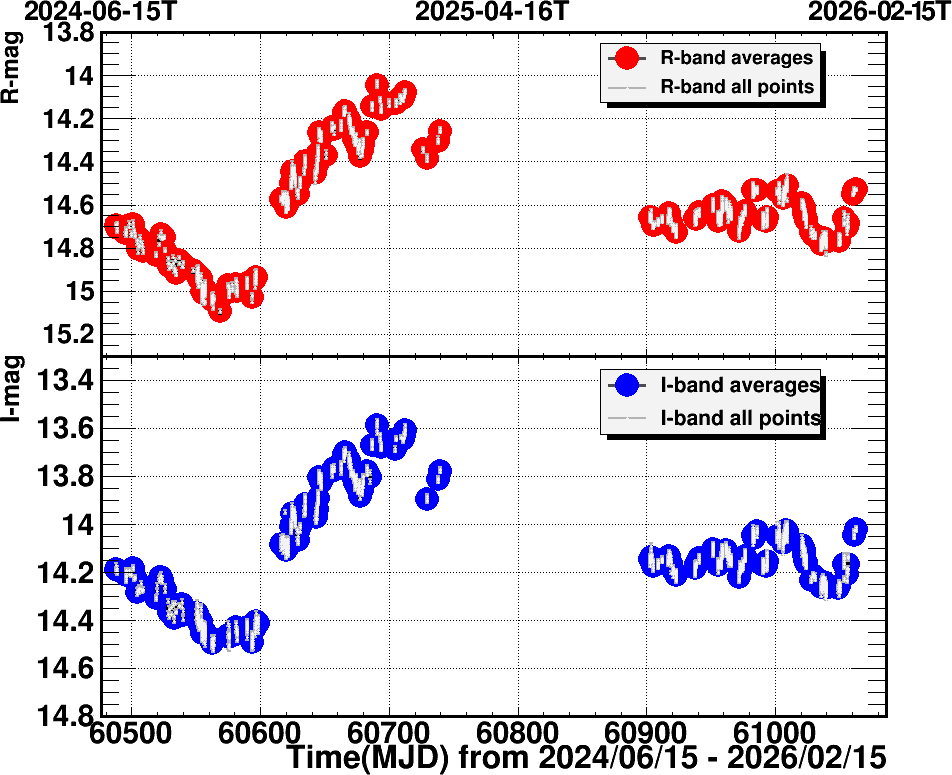}

\vspace{0.1cm}
\textbf{1ES~1959+650}

\vspace{0.4cm}

\includegraphics[width=\textwidth,height=0.28\textheight,keepaspectratio]{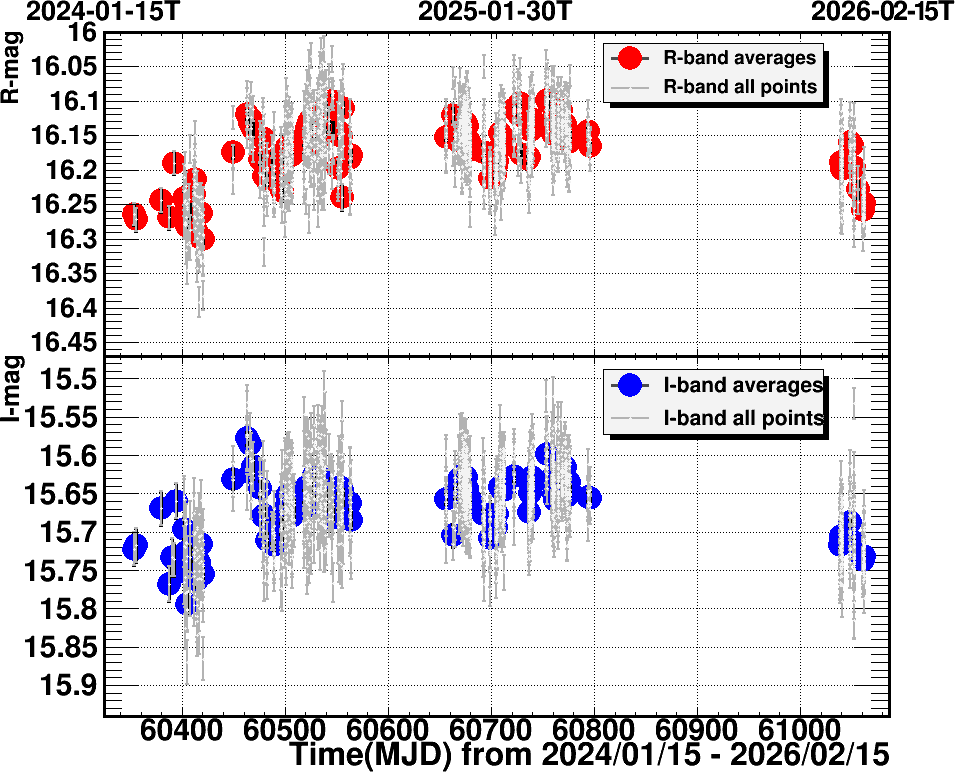}

\vspace{0.1cm}
\textbf{H~1426+428}

\vspace{0.1cm}

\includegraphics[width=\textwidth,height=0.28\textheight,keepaspectratio]{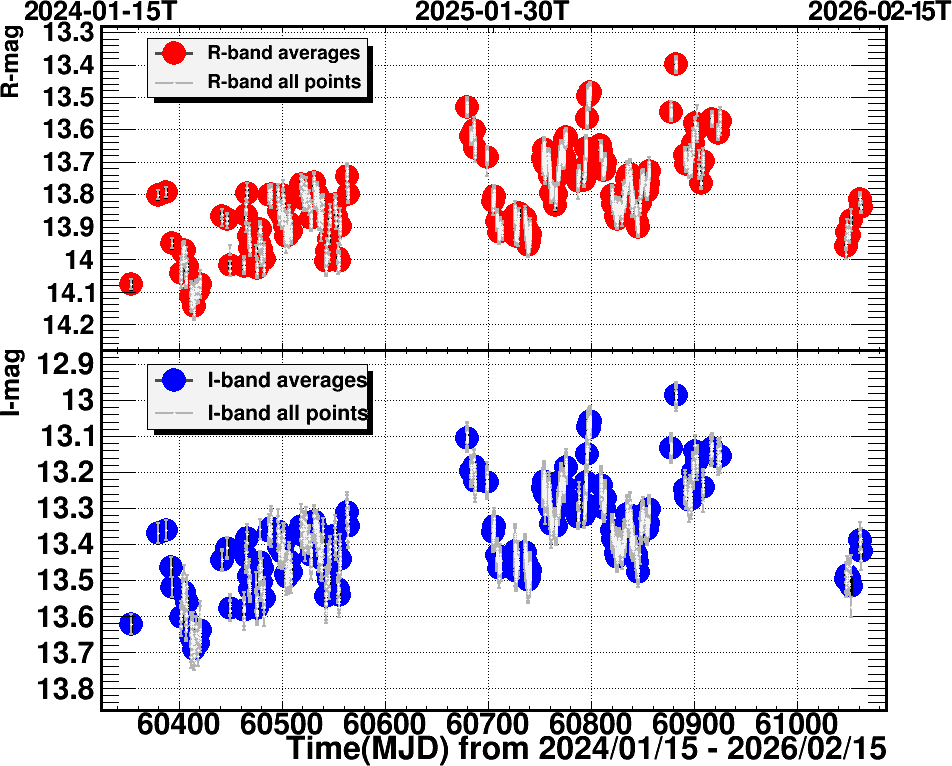}

\vspace{0.2cm}
\textbf{PG~1553+113}

\caption{Long-term optical light curves of the three monitored blazars in the $R$ (top panels) and $I$ (bottom panels) bands during the 2024--2026 campaign. Grey symbols represent individual exposures, while larger coloured symbols indicate nightly averaged magnitudes. Error bars correspond to photometric uncertainties, typically smaller than the symbol size.}
\label{fig:lt_all}

\end{figure*}

To illustrate both the temporal coverage and the long-term
optical evolution of the targets, Figure~\ref{fig:lt_all}
shows the complete set of individual photometric measurements
obtained during the 2024--2026 monitoring campaign.

Grey points represent single-exposure magnitudes,
highlighting the dense and regular seasonal sampling achieved
over the full biennium. Nightly averaged magnitudes
are overplotted with larger symbols to provide a clearer
representation of the long-term variability trends.

Error bars correspond to the photometric uncertainties
associated with each exposure. Owing to the high
signal-to-noise ratio of the dataset, these uncertainties are
typically smaller than the symbol size.

The dense distribution of measurements demonstrates that
the three sources were monitored regularly throughout the
entire observing period, rather than during isolated time
intervals. Consequently, the derived intranight duty cycles are not dominated
by a small number of isolated observing intervals.

Apparent band-like structures visible in some epochs
(particularly in the case of H~1426+428) are produced by
night-to-night zero-point offsets affecting all objects in the
field. These common-mode variations are verified through
the behaviour of the comparison stars and do not influence
the intranight differential analysis, which is performed
independently on a night-by-night basis using blazar--star
and star--star light curves (Sect.~4).

No comparison-star light curves are shown in these long-term
panels, as their stability is demonstrated through the
intranight star--star differential curves discussed in
Sect.~3.2.

The peak-to-peak amplitudes derived from the nightly
averaged magnitudes indicate that 1ES~1959+650 is the most
variable source in the sample, with long-term variations of
about 1.05 mag in $R$ and 0.91 mag in $I$.

H~1426+428 shows comparatively small oscillations,
of only $\sim$0.2 mag in both bands, whereas PG~1553+113
exhibits intermediate variability, with amplitudes of
approximately 0.75 mag in $R$ and 0.70 mag in $I$.

\subsection{Intranight variability}

The search for IDV was carried out on the well-sampled
observing sessions selected according to the criterion defined
in Sect.~4. The classical F-test was applied independently
using the two comparison-star pairs defined in Sect.~4.2 for
each source.

Following the classification scheme described in Sect.~4.2,
a night was classified as V only when both pair-level F
statistics exceeded the critical value corresponding to
$\alpha=0.001$. Sessions in which one comparison-star pair
satisfied the V condition and the other satisfied the
$\alpha=0.01$ condition were classified as PV, while all
remaining cases were classified as NV.

\subsubsection{1ES~1959+650}

For 1ES~1959+650, 49 nights in the $R$ band and
48 nights in the $I$ band were analysed.
None of these sessions fulfils the conditions required to be
classified as V or PV. All analysed sessions are therefore
classified as NV in both filters
(Table~\ref{tab:idv_summary}).

Within the sensitivity limits of the dataset and the adopted
classification criteria, no statistically significant IDV is
detected in 1ES~1959+650.

\subsubsection{H~1426+428}

H~1426+428 is the only source in the sample showing
statistically robust IDV.
Among the analysed sessions, 15 out of 43 nights in the
$R$ band and 5 out of 40 nights in the $I$ band satisfy
the most stringent criterion for classification as V.
In addition, 9 nights in the $R$ band and 11 nights in the
$I$ band are classified as PV
(Table~\ref{tab:idv_summary}).

\begin{figure*}
\centering

\subfloat[$R$ band, 2025 February 5--6]{
\includegraphics[width=0.48\textwidth]{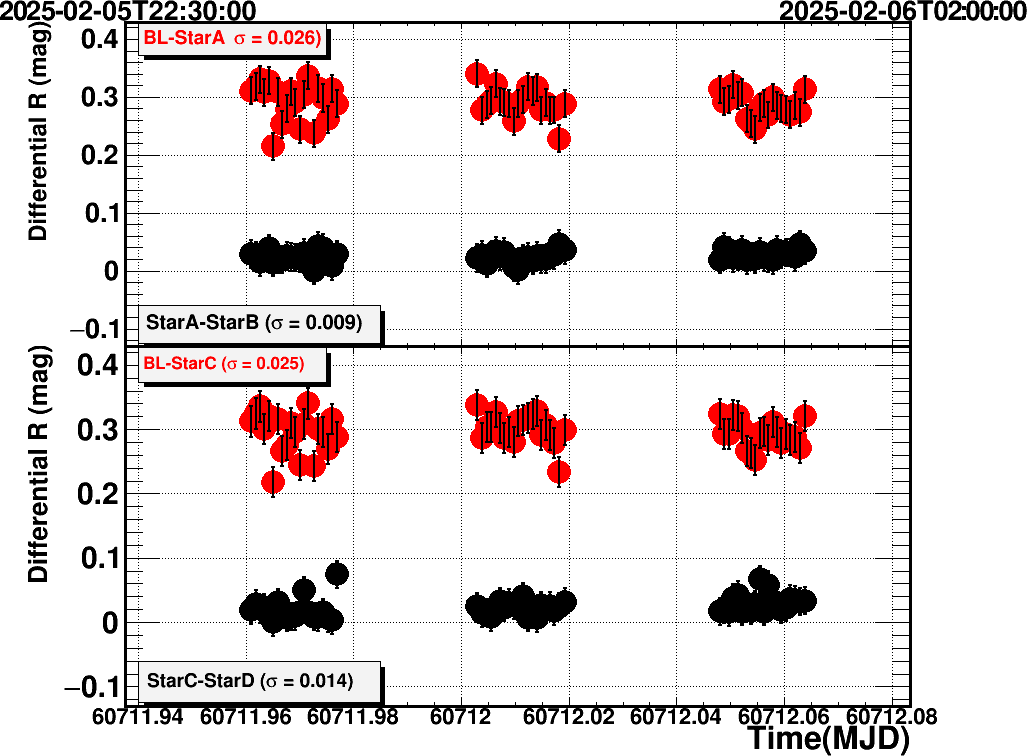}
}
\hfill
\subfloat[$I$ band, 2026 January 12]{
\includegraphics[width=0.48\textwidth]{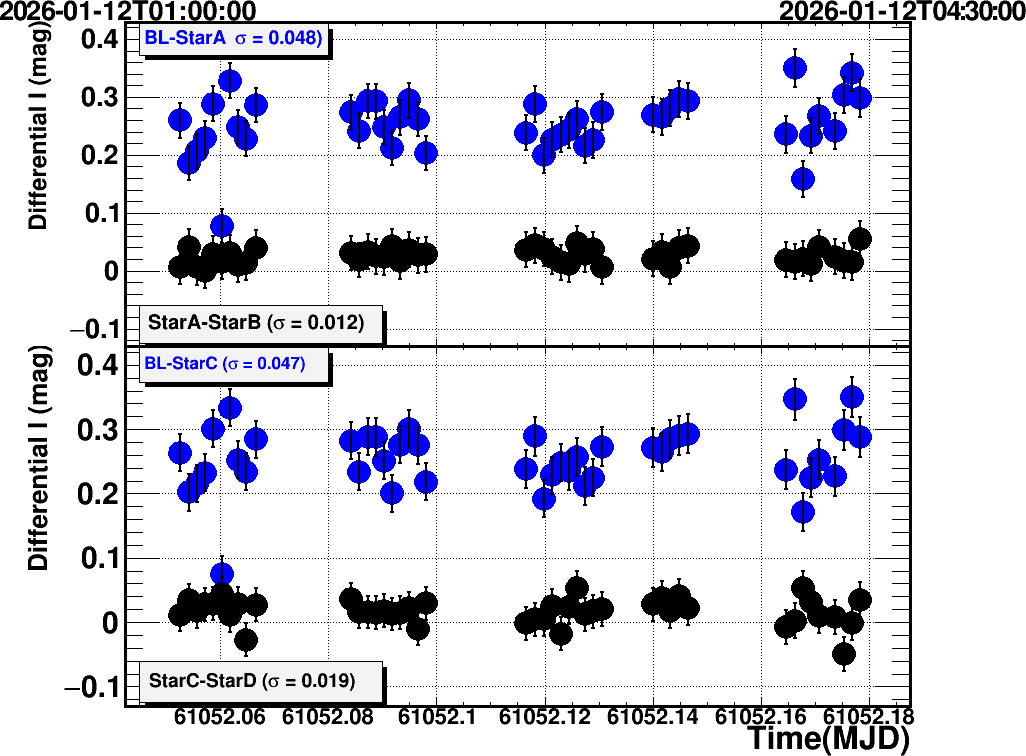}
}

\caption{
Representative examples of nights classified as variable (V) for
H~1426+428 in the $R$ and $I$ bands.
The left panel shows the $R$-band light curves on 2025 February 5--6,
while the right panel shows the $I$-band light curves on 2026 January 12.
In each panel, the upper subpanel shows the differential light curves
between the blazar and one comparison star together with the corresponding
star--star differential curve. The lower subpanel shows the same quantities
for the second independent comparison-star pair.
For H~1426+428, Star~A, Star~B, Star~C, and Star~D correspond
to the generic labels $S_1$, $S_2$, $S_3$, and $S_4$ used in
the variability analysis. For clarity, the differential magnitudes
are vertically shifted by constant offsets. The $\sigma$ values
reported in the labels denote the rms scatter of the corresponding
differential light curves. In both filters, the low scatter of the star--star light
curves supports the interpretation that the observed variations
are intrinsic to the blazar and not caused by photometric
instabilities.}
\label{fig:idv_1426_example}
\end{figure*}

The corresponding intranight variability amplitudes for
V nights are typically at the level of
$\sim10$--$15\%$.
Figure~\ref{fig:idv_1426_example} shows representative
examples of nights classified as V in the $R$ and $I$ bands.
In both filters and for both independent comparison-star
pairs, the blazar--star differential light curves exhibit
clear variations, while the star--star differential curves
remain stable, supporting an intrinsic origin of the observed
flux changes.

\subsubsection{PG~1553+113}

For PG~1553+113, 70 nights in the $R$ band and
66 nights in the $I$ band were analysed.
No observing session satisfies the adopted variability
conditions required for classification as V or PV.
All nights are therefore classified as NV in both filters
(Table~\ref{tab:idv_summary}).

Thus, despite the large number of well-sampled observing
sessions, PG~1553+113 does not exhibit statistically
significant IDV within the sensitivity limits of the present
monitoring campaign. This non-detection should be understood
within the adopted statistical framework and does not exclude
the possibility of rare, lower-amplitude, or state-dependent
IDV episodes below the sensitivity of the present dataset.

\subsubsection{Robustness checks}

As a robustness check, we repeated the F-test classification by
changing the independent pairing of the four comparison stars.
In addition to the adopted configuration,
$(S_1,S_2)+(S_3,S_4)$, we tested the alternative pairings
$(S_1,S_3)+(S_2,S_4)$ and $(S_1,S_4)+(S_2,S_3)$.
These checks were used only to assess the sensitivity of the
classification to the adopted comparison-star pairing; the
classification reported in Table~\ref{tab:idv_summary} is based
on the pair definition described in Sect.~4.2.

For PG~1553+113, all sessions remain classified as NV in
both filters for all pairing choices. This confirms that the absence of detected IDV in this source is not driven by the adopted comparison-star pairing, including
the pairing involving the fainter comparison stars.
For H~1426+428, the number of V sessions remains broadly stable,
especially in the $R$ band, where the three pairings give
15, 15, and 13 V nights, respectively. In the $I$ band, the
corresponding numbers are 5, 4, and 4. This supports the main
result that H~1426+428 is the only source in the sample showing
secure IDV, while also reinforcing the caution adopted in this
work against overinterpreting the difference between the
$R$- and $I$-band duty cycles.

We also performed a simple Monte Carlo null test to assess
whether the observed incidence of V classifications in
H~1426+428 could be produced by photometric noise alone.
For each observing session, the blazar light curve was replaced
by a non-variable synthetic source with the same mean magnitude,
time sampling, and photometric uncertainties as the observed
blazar data. The same F-test classification procedure was then
applied to the simulated light curves. The expected number of
false V classifications is well below one session per filter,
whereas the observed numbers are 15 V nights in the $R$ band
and 5 in the $I$ band. This supports the conclusion that the
IDV detected in H~1426+428 is not an artefact of the
photometric noise level alone.

\subsection{Duty cycle of intranight variability}

We quantified the incidence of IDV using a duration-weighted
duty cycle, defined for each source and filter as the fraction
of the total monitoring time associated with observing sessions
classified as V and for which a local characteristic variability
timescale could be measured. The quantity is defined as

\begin{equation}
DC_{\rm dur} =
100 \times
\frac{\sum_{i=1}^{n} V_{\tau,i}\,\Delta t_i}
{\sum_{i=1}^{n} \Delta t_i},
\end{equation}

where $n$ is the number of analysed observing sessions for a
given source and filter, and $\Delta t_i$ is the duration of the
$i$-th monitoring session. The quantity $V_{\tau,i}$ is set to
1 if the session is classified as V and contains at least one
consecutive flux variation significant at the adopted $3\sigma$
level, from which a local characteristic timescale can be
estimated. In all other cases, including PV sessions and V
sessions without a qualifying consecutive variation,
$V_{\tau,i}=0$. The number of sessions satisfying this condition
is reported as $V_{\tau}$ in Table~\ref{tab:idv_summary}.

With this definition, we obtain
$DC_{{\rm dur},R}=28.0\%$ and $DC_{{\rm dur},I}=7.0\%$
for H~1426+428. No statistically significant IDV was detected
for 1ES~1959+650 or PG~1553+113, resulting in
$DC_{\rm dur}=0\%$ for both sources and filters
(Table~\ref{tab:idv_summary}). Since this duration-weighted
definition is restricted to V sessions with a measured local
timescale, direct numerical comparisons with duty cycles
computed using different definitions should be made with
caution.

The derived duration-weighted duty cycle for H~1426+428
indicates that optical IDV in this source is intermittent and
occurs only during a fraction of the monitored time. We do not
interpret the difference between the $R$- and $I$-band duty
cycles as firm evidence for wavelength-dependent variability,
since several epochs in the $I$ band satisfy only the PV
criterion under the adopted classification scheme.

For H~1426+428, local characteristic variability timescales
were estimated only for V sessions showing significant
consecutive magnitude changes at the $3\sigma$ level. If more
than one consecutive pair satisfied this criterion within the
same session, the shortest value was retained as the most rapid
local rate of variation. These values should be interpreted as
local extrapolated doubling/halving timescales, rather than as
direct measurements of fully resolved flux-doubling or
flux-halving events. The resulting timescales, reported in
Table~\ref{tab:tau_1426}, range between approximately
5 and 15~min in the observer frame.

The timescales were estimated assuming exponential flux
variations, following the doubling/halving prescription commonly
adopted in blazar variability studies
\citep[e.g.][]{Valtaoja1999,Foschini2011}. For two consecutive
measurements at times $t_0$ and $t$, the observer-frame
timescale was computed as

\begin{equation}
|\tau_{\rm obs}| =
\frac{|t-t_0|\,\ln 2}
{|\ln [F(t)/F(t_0)]|},
\end{equation}

with $F(t)/F(t_0)=10^{-0.4\Delta m}$. A first-order estimate
of the uncertainty was obtained by propagating the uncertainty
on the consecutive magnitude difference and neglecting the
timing uncertainty. Since $\tau \propto 1/|\Delta m|$, this gives

\begin{equation}
\frac{\sigma_{\tau}}{\tau} \simeq \frac{1}{S},
\end{equation}

where $S$ is the significance of the consecutive magnitude
variation reported in Table~\ref{tab:tau_1426}. For the values
listed in the table, this corresponds to relative uncertainties of
approximately 12--32\%, or about 0.7--3.8~min in
$|\tau_{\rm obs}|$.

The observer-frame timescale was corrected for cosmological
time dilation as

\begin{equation}
\tau_{\rm int} = \frac{\tau_{\rm obs}}{1+z},
\end{equation}

using $z=0.129$ for H~1426+428.

\begin{table*}
\centering
\caption{
Summary of the intranight variability results for the three
monitored blazars.
Only well-sampled observing sessions satisfying the selection
criterion defined in Sect.~4 were considered.
A session is classified as V if both pair-level F-test statistics
exceed the critical value corresponding to $\alpha=0.001$.
PV indicates probably variable sessions and NV non-variable
sessions.
The column $V_{\tau}$ reports the subset of V sessions showing
significant consecutive flux variations from which a local
characteristic variability timescale could be estimated.
The last column reports the duration-weighted duty cycle,
computed as the fraction of the total monitoring time associated
with $V_{\tau}$ sessions.
}
\label{tab:idv_summary}
\begin{tabular}{lccccccc}
\hline\hline
Source & Filter & Sessions & V & $V_\tau$ & PV & NV & $DC_{\rm dur}$ (\%) \\
\hline
1ES~1959+650 & R & 49 & 0 & 0 & 0 & 49 & 0.0 \\
1ES~1959+650 & I & 48 & 0 & 0 & 0 & 48 & 0.0 \\
H~1426+428   & R & 43 & 15 & 13 & 9 & 19 & 28.0 \\
H~1426+428   & I & 40 & 5 & 3 & 11 & 24 & 7.0 \\
PG~1553+113  & R & 70 & 0 & 0 & 0 & 70 & 0.0 \\
PG~1553+113  & I & 66 & 0 & 0 & 0 & 66 & 0.0 \\
\hline
\end{tabular}
\end{table*}

\begin{table*}
\centering
\caption{
Characteristic intranight variability timescales measured for
H~1426+428.
}
\label{tab:tau_1426}
\setlength{\tabcolsep}{4pt}
\begin{tabular}{lcccccccccc}
\hline\hline
Source & Filter & Night & $t_0$ & $t$ & $\Delta m$ &
$\Delta F/F_0$ (\%) & Signif. $S$ & $|\tau|$ (min) &
$|\tau_{\rm int}|$ (min) & R/D \\
\hline
H~1426+428 & R & 2024-08-09 & 60531.96545 & 60531.96620 & 0.1514 & -13.01 & 5.81 & 5.41 & 4.79 & D \\
H~1426+428 & I & 2024-12-28 & 60672.20355 & 60672.20451 & -0.0984 & 9.49 & 3.13 & 10.55 & 9.34 & R \\
H~1426+428 & R & 2025-01-01 & 60676.10800 & 60676.10881 & 0.0733 & -6.52 & 3.18 & 11.97 & 10.60 & D \\
H~1426+428 & R & 2025-01-02 & 60677.07200 & 60677.07281 & -0.0987 & 9.52 & 3.90 & 8.88 & 7.87 & R \\
H~1426+428 & R & 2025-01-04 & 60679.11946 & 60679.12027 & 0.0778 & -6.92 & 3.31 & 11.27 & 9.98 & D \\
H~1426+428 & R & 2025-02-04 & 60710.98104 & 60710.98185 & 0.0895 & -7.91 & 3.88 & 9.78 & 8.66 & D \\
H~1426+428 & R & 2025-02-05 & 60711.96434 & 60711.96515 & -0.0977 & 9.41 & 4.21 & 8.95 & 7.93 & R \\
H~1426+428 & R & 2025-02-06 & 60712.96682 & 60712.96765 & 0.0973 & -8.57 & 4.13 & 9.27 & 8.21 & D \\
H~1426+428 & R & 2025-02-15 & 60721.99472 & 60721.99553 & -0.1043 & 10.09 & 4.52 & 8.39 & 7.43 & R \\
H~1426+428 & R & 2025-03-05 & 60739.88525 & 60739.88606 & 0.0915 & -8.09 & 3.95 & 9.55 & 8.46 & D \\
H~1426+428 & R & 2025-03-25 & 60759.87364 & 60759.87445 & -0.1309 & 12.81 & 5.65 & 6.69 & 5.92 & R \\
H~1426+428 & R & 2025-04-08 & 60773.86325 & 60773.86476 & 0.1093 & -9.58 & 5.12 & 14.90 & 13.20 & D \\
H~1426+428 & I & 2025-12-31 & 61040.18488 & 61040.18640 & 0.1183 & -10.32 & 3.82 & 13.89 & 12.31 & D \\
H~1426+428 & R & 2026-01-01 & 61041.08491 & 61041.08641 & -0.1110 & 10.77 & 4.53 & 14.67 & 12.99 & R \\
H~1426+428 & I & 2026-01-12 & 61052.06030 & 61052.06181 & 0.2593 & -21.24 & 8.68 & 6.28 & 5.56 & D \\
H~1426+428 & R & 2026-01-12 & 61052.11037 & 61052.11188 & -0.1221 & 11.91 & 4.78 & 13.34 & 11.81 & R \\
\hline
\end{tabular}
\tablefoot{
Only sessions classified as V and with a measured local
characteristic timescale are reported.
Times $t_0$ and $t$ are given in MJD and refer to the two
consecutive measurements used to estimate the timescale.
The quantity $\Delta m$ is defined as $m(t)-m(t_0)$, while
the corresponding relative flux variation is computed as
$\Delta F/F_0 = [F(t)-F(t_0)]/F(t_0)
=10^{-0.4\Delta m}-1$, with $F_0=F(t_0)$, and is expressed
in percent.
The significance $S$ is reported in units of $\sigma$.
The absolute values of the observed characteristic timescale
$|\tau|$ and of the intrinsic timescale
$|\tau_{\rm int}|=|\tau|/(1+z)$ are given in minutes.
If more than one consecutive pair satisfies the adopted
$3\sigma$ criterion within the same session, the shortest
$|\tau|$ value is reported.
These timescales represent local extrapolated doubling/halving
times inferred from consecutive flux changes and do not imply
that a full flux doubling or halving was observed within a
single night.
Neglecting timing uncertainties, a first-order estimate gives
$\sigma_{\tau}/\tau \simeq 1/S$; for the entries reported here
this corresponds to relative uncertainties of approximately
12--32\%, or about 0.7--3.8~min in $|\tau|$.
The last column indicates a flux rise (R, $\Delta m<0$) or
decay (D, $\Delta m>0$).
}
\end{table*}

\subsection{Distribution of $F$-values}

\begin{figure*}
\centering

\subfloat[H~1426+428]{
\includegraphics[width=0.32\textwidth]{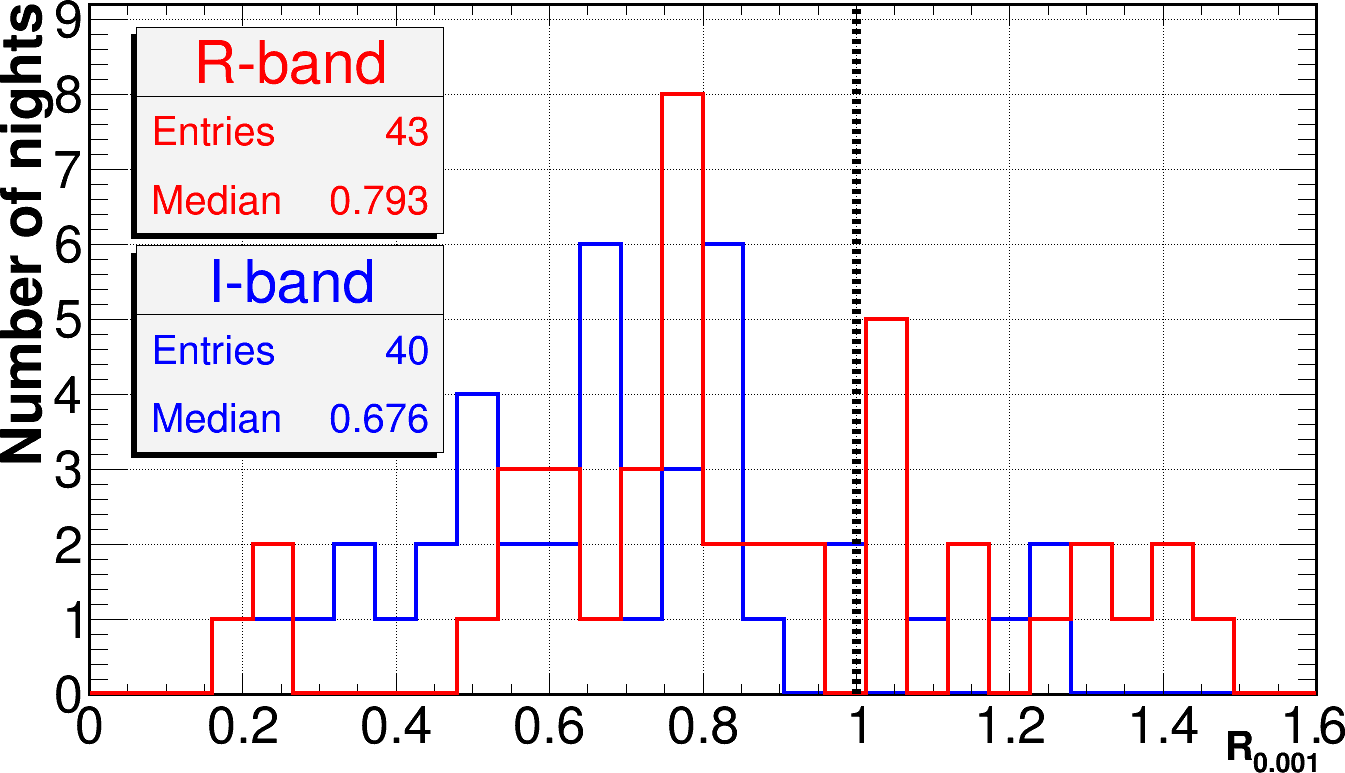}
}
\hfill
\subfloat[1ES~1959+650]{
\includegraphics[width=0.32\textwidth]{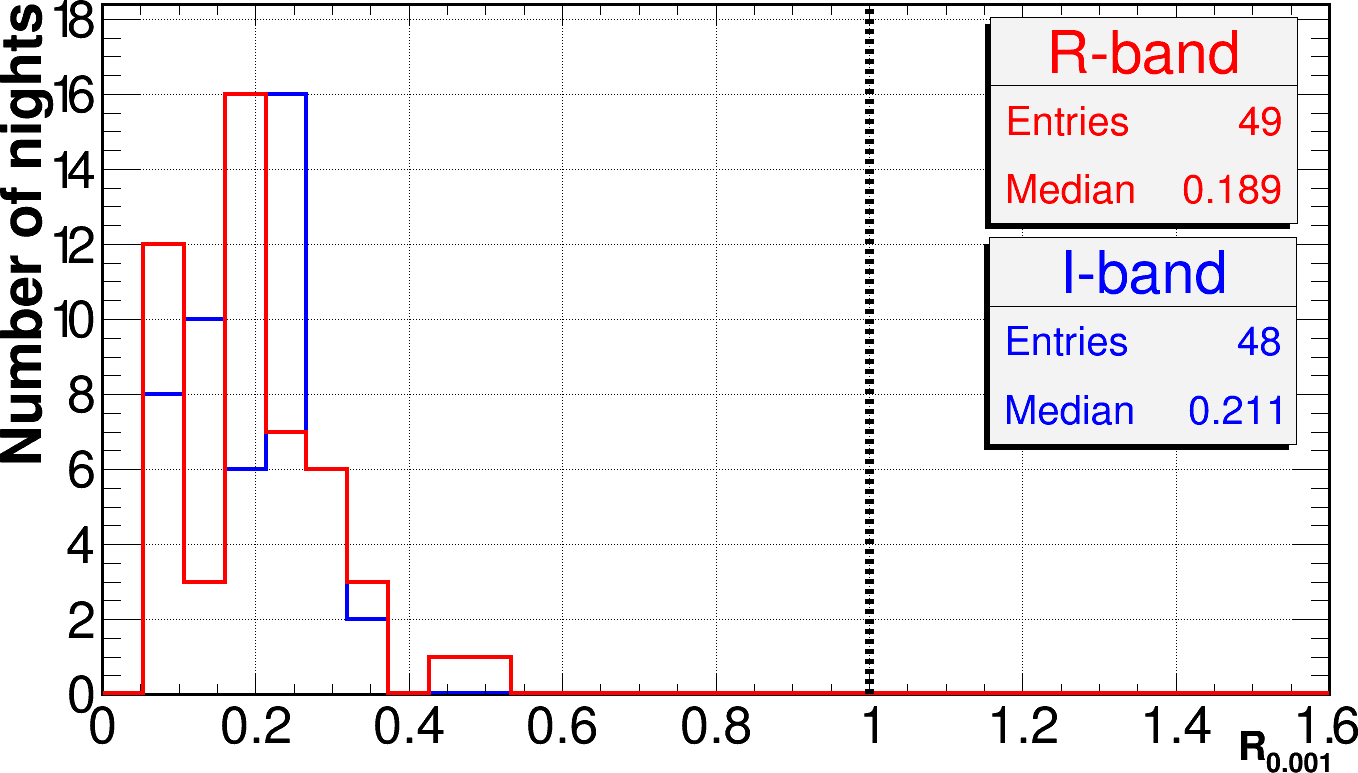}
}
\hfill
\subfloat[PG~1553+113]{
\includegraphics[width=0.32\textwidth]{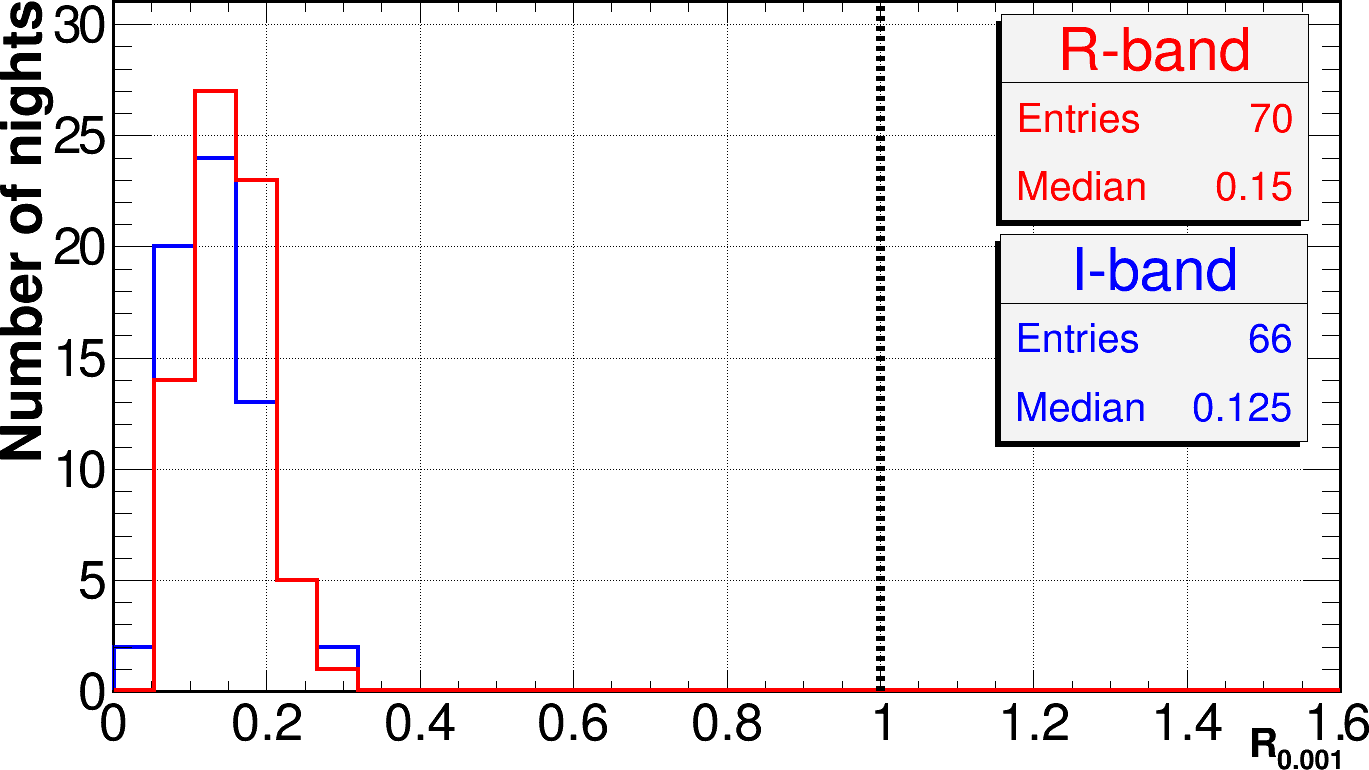}
}

\caption{
Distribution of the normalized F-test statistic for all analysed
nights with $N \geq 20$ photometric measurements. The plotted
quantity is
$R_{0.001}=\min(F_{12}/F_{{\rm crit},0.001},
F_{34}/F_{{\rm crit},0.001})$.
The vertical dashed line marks the variability threshold
($R_{0.001}=1$), corresponding to a significance level
$\alpha=0.001$.
Red and blue histograms represent the distributions obtained
in the $R$ and $I$ bands, respectively.
Only H~1426+428 exhibits nights above the adopted threshold,
whereas the distributions for 1ES~1959+650 and PG~1553+113
remain entirely below $R_{0.001}=1$.
}

\label{fig:f_distribution}

\end{figure*}

Figure~\ref{fig:f_distribution} shows the distribution of the
normalized F-test statistic $R_{0.001}$ for all analysed sessions
in the three sources,

\begin{equation}
R_{0.001}=
\min\left(
\frac{F_{12}}{F_{{\rm crit},0.001}},
\frac{F_{34}}{F_{{\rm crit},0.001}}
\right),
\end{equation}

where $F_{{\rm crit},0.001}$ is the critical value corresponding
to $\alpha=0.001$ and $\nu=N-1$ degrees of freedom.
Here $F_{12}$ and $F_{34}$ are the pair-level F-test statistics
defined in Sect.~4.2 for the two independent comparison-star
pairs $(S_1,S_2)$ and $(S_3,S_4)$, respectively.
With this definition, the variability threshold corresponds to
$R_{0.001}=1$, requiring simultaneous significance for both
comparison-star pairs.

For H~1426+428, the $R$-band distribution
(43 nights) has a median
$R_{0.001}=0.793$ with a 95th percentile of 1.403,
while the $I$-band distribution (40 nights) has a
median of 0.676 and a 95th percentile of 1.234.
Several nights exceed the adopted variability threshold
in both filters, confirming the presence of statistically
robust intranight variability in this source.

The relatively high median value of the normalized
statistic in the $R$ band for H~1426+428 reflects the
presence of several nights with enhanced variance in the
blazar differential light curves, although only a subset
of them exceeds the adopted variability threshold.
This behaviour is expected for sources exhibiting
intermittent intranight activity close to the adopted
significance limit.

In contrast, both 1ES~1959+650 and PG~1553+113 show
distributions entirely below the critical threshold in
both filters.
For 1ES~1959+650, the medians are 0.189
($R$ band, 49 nights) and 0.211
($I$ band, 48 nights), with corresponding 95th
percentiles of 0.357 and 0.296, respectively.
Similarly, for PG~1553+113 the medians are 0.150
($R$ band, 70 nights) and 0.125
($I$ band, 66 nights), with 95th percentiles of
0.231 and 0.230, respectively.
No night in these two sources satisfies the condition
$R_{0.001}>1$.

The distributions are therefore strongly peaked below
unity for 1ES~1959+650 and PG~1553+113, whereas
H~1426+428 exhibits a significant high-ratio tail,
reflecting its higher incidence of intranight variability.

This behaviour indicates that the observed scatter of the
blazar differential light curves is generally comparable
to the photometric noise level traced by the star--star
curves.
Only in the case of H~1426+428 does the distribution
develop a significant high-ratio tail, revealing genuine
intranight flux variability.

\section{Discussion}

The optical monitoring campaign presented in this work provides
constraints on the intranight variability properties of three
TeV-emitting HBLs: 1ES~1959+650, H~1426+428, and
PG~1553+113. The use of a uniform dataset, obtained with the
same telescope, instrumental configuration, reduction procedure,
and statistical classification scheme, allows the three sources
to be compared within the same observational and statistical
framework.

Among the analysed targets, statistically significant IDV is
detected only in H~1426+428. In this source, 15 out of
43 analysed sessions in the $R$ band and 5 out of 40 sessions
in the $I$ band satisfy the most stringent variability criterion.
Local characteristic timescales could be measured for 13 of the
V sessions in the $R$ band and for 3 in the $I$ band, leading
to duration-weighted duty cycles of
$DC_{{\rm dur},R}=28.0\%$ and
$DC_{{\rm dur},I}=7.0\%$, respectively.
These values indicate that rapid optical synchrotron variability
in H~1426+428 is intermittent rather than continuously present
during the monitoring campaign.

The detection of IDV in H~1426+428 is consistent with recent
optical studies of this source. \citet{Chang2023} reported
intraday variability in H~1426+428 during high time-resolution
multicolour monitoring, including a candidate quasi-periodic
oscillation with a period of about 49~min. More recently,
\citet{Chang2024} detected IDV on seven nights and discussed
the variability in terms of turbulence in the relativistic jet
and magnetic reconnection processes. Our results extend this
picture using a larger and internally consistent dataset, showing
that H~1426+428 is the only source in the present sample with
secure optical IDV detections, while also indicating that this
activity is episodic.

The robustness checks performed in this work support the
reliability of this result. Changing the independent pairing of
the comparison stars leaves PG~1553+113 classified as
non-variable in both filters and gives broadly consistent results
for H~1426+428, especially in the $R$ band. In addition, a
simple Monte Carlo null test shows that the expected number of
false V classifications due to photometric noise alone is well
below one session per filter, whereas the observed number of V
sessions in H~1426+428 is substantially larger. These checks
indicate that the main IDV result is not driven by the adopted
comparison-star pairing or by the photometric noise level alone.

The inferred timescales, ranging from approximately 5 to
15~min in the observer frame, are short for optical IDV and
suggest that, during the detected events, a compact and rapidly
varying optical component contributes to the observed emission.
In a high-energy blazar context, such a component can be
interpreted as a localized enhancement of the synchrotron
emission from the relativistic jet. Possible physical origins
include localized particle acceleration, turbulence, shock
propagation, or magnetic reconnection. The presence of both
rising and decaying events is consistent with short-lived
enhancements followed by fading of the synchrotron emission.

These timescales should, however, be interpreted with caution.
They are local extrapolated doubling/halving timescales inferred
from significant consecutive flux changes, rather than direct
measurements of fully resolved flare rise or decay durations.
The first-order uncertainty estimate discussed in Sect.~5.3
indicates relative uncertainties of about 12--32\% for the
reported values. The measured timescales are therefore best
regarded as characteristic local rates of optical flux variation,
not as complete flare durations.

No robust optical IDV is detected in 1ES~1959+650. This result
is consistent with several previous optical monitoring studies of
this source. \citet{Gaur2012} did not detect genuine
microvariability in 1ES~1959+650 during quasi-simultaneous
two-band monitoring, while \citet{Zhang2017} found no
intranight variability in 38 light curves collected over 19 nights,
despite detecting significant short- and long-term variability.
On the other hand, \citet{Yuan2015} reported intraday variability
during two nights in September 2009. Taken together, these
results suggest that optical IDV in 1ES~1959+650, if present,
is rare and likely connected to specific activity states.

A similar conclusion applies to PG~1553+113. In the present
dataset, all 70 analysed sessions in the $R$ band and all
66 analysed sessions in the $I$ band are classified as NV.
The additional test based on alternative comparison-star pairings
confirms that this non-detection is not caused by the adopted
comparison-star pairing, including the pairing involving the
fainter comparison stars. Previous studies have shown that
PG~1553+113 can display optical IDV, but with a relatively low
duty cycle. For example, \citet{Agarwal2021} reported an
intranight duty cycle of about 10--18\%, while
\citet{Dhiman2023}, using the power-enhanced F-test and nested
ANOVA, detected significant variability only during a small
fraction of the monitored nights. The absence of robust IDV in
our campaign is therefore not in contradiction with the literature.
It indicates that high-confidence optical IDV in PG~1553+113 is
not frequent in the monitored states, at least within the sensitivity
and classification criteria adopted here.

The different behaviour observed among the three sources is
consistent with the broader picture of optical variability in
high-synchrotron-peaked BL Lac objects. HBLs generally show
lower optical intranight duty cycles than low-synchrotron-peaked
blazars. \citet{Paliya2017} found that high-synchrotron-peaked
objects exhibit substantially lower optical intranight duty cycles
than low-synchrotron-peaked (LSP) blazars. Similarly,
\citet{Goyal2013} showed that TeV blazars display variability
properties distinct from those of other AGN classes. The present
results are consistent with this picture: robust optical IDV is
detected in only one of the three TeV HBLs, and even in that case
the activity is intermittent.

This behaviour can be understood in terms of the location of the
synchrotron peak in HBLs. Since the synchrotron peak is located
at UV or X-ray frequencies, the optical band samples a lower-energy
part of the synchrotron component. Rapid variability may therefore
be more prominent at higher energies, while optical IDV may become
detectable only when compact and rapidly varying synchrotron
components contribute significantly to the optical flux. This
interpretation is consistent with the absence of frequent optical
IDV in 1ES~1959+650 and PG~1553+113 despite their known
long-term variability.

From a high-energy perspective, the present results show that
rapid optical synchrotron variability is not continuously present
in TeV HBLs, even in sources known to be variable at X-ray and
TeV energies. This has practical implications for coordinated
multiwavelength campaigns. Optical monitoring can provide dense
and relatively inexpensive temporal coverage of the low-energy
synchrotron component, but the absence of optical IDV during most
observing sessions indicates that rapid optical activity is strongly
state-dependent. Optical follow-up of TeV HBLs is therefore most
informative when combined with high-energy alerts or with
contemporaneous X-ray and gamma-ray observations.

The statistical methodology adopted in this work was designed to
favour reliability over sensitivity. The classical F-test has lower
statistical power than ANOVA under ideal conditions, as discussed
by \citet{deDiego2010}, but it remains a valid variance-based test
when applied consistently to differential light curves. In the present
monitoring programme, the alternation of $R$- and $I$-band exposure
blocks produces non-uniform temporal segmentation in individual
filters, making a systematic application of ANOVA non-homogeneous
across all nights and sources. For this reason, we adopted the same
F-test framework for the entire dataset, using two independent
comparison-star pairs and requiring agreement between them for a
secure variability detection.

This restrictive approach reduces the probability of spurious
detections caused by individual comparison-star instabilities or
photometric systematics. It also explains why the duty cycles
reported here may be lower than those obtained with less restrictive
definitions: marginal or single-pair detections are not counted as
secure variability events. Since our duty cycle is duration-weighted
and restricted to V sessions with a measured local characteristic
timescale, direct numerical comparisons with duty cycles computed
using different definitions should be made with caution. The values
reported here should therefore be interpreted as conservative
estimates of high-confidence optical IDV activity.

Overall, the results indicate that optical IDV in TeV HBLs is
episodic and state-dependent. H~1426+428 shows secure optical
IDV and measurable local characteristic timescales, whereas
1ES~1959+650 and PG~1553+113 remain non-variable within the
sensitivity and classification criteria adopted here. Thus, strong
long-term optical variability or established TeV emission does not
necessarily imply frequent hour-scale optical variability.

Long-term uniform monitoring remains essential for constraining
the duty cycle of optical IDV and for distinguishing rare intrinsic
variability events from statistical or instrumental artefacts. The
results obtained in the present work have motivated the observing
strategy adopted for the continuation of the monitoring programme
during the 2026--2028 biennium with the Asiago Schmidt telescope.
H~1426+428 will remain a primary target for dedicated intranight
monitoring sessions, owing to its confirmed optical IDV activity
and measurable local characteristic timescales. For 1ES~1959+650
and PG~1553+113, intensive intranight observations will be most
useful during periods of enhanced activity or in response to
external high-energy alerts, while long-term optical monitoring
will continue to provide the baseline needed to identify such states.

\section{Conclusions}

We presented a two-year homogeneous optical monitoring campaign
of three TeV-emitting high-synchrotron-peaked BL Lac objects,
1ES~1959+650, H~1426+428, and PG~1553+113, carried out with
the Asiago Schmidt 67/92~cm telescope between February 2024
and January 2026.

The dataset consists of more than 17\,000 photometric measurements
obtained in the $R$ and $I$ bands over 957 source--filter observing
sessions, corresponding to a total monitoring time of about
420~h. The observations provide an internally consistent dataset
for investigating optical IDV in TeV HBLs and for comparing
detections and non-detections under the same statistical criteria.
In the context of high-energy time-domain astrophysics, this
dataset provides a dense optical view of the synchrotron component
of relativistic jets on timescales that are difficult to sample
continuously with X-ray and gamma-ray facilities.

The IDV search was performed using a classical F-test approach
designed to favour reliability over sensitivity. For each source,
two independent pairs of comparison stars were used, and a session
was classified as V only when both comparison-star pairs satisfied
the most stringent variability criterion. Only well-sampled sessions
with $N\geq20$ photometric measurements per filter were included
in the IDV analysis.

The main results can be summarised as follows:

\begin{itemize}

\item H~1426+428 is the only source in the sample showing
statistically secure optical IDV. In this source, 15 out of
43 analysed sessions in the $R$ band and 5 out of 40 sessions
in the $I$ band are classified as V.

\item Local characteristic variability timescales were measured
for 13 V sessions in the $R$ band and for 3 in the $I$ band.
Using only these events for the duration-weighted duty-cycle
computation, we obtain
$DC_{{\rm dur},R}=28.0\%$ and
$DC_{{\rm dur},I}=7.0\%$ for H~1426+428.

\item The measured timescales range from approximately
5 to 15~min in the observer frame. They should be interpreted
as local extrapolated doubling/halving timescales inferred from
significant consecutive flux changes, rather than as direct
measurements of complete flare durations.

\item No robust optical IDV is detected in 1ES~1959+650.
All 49 analysed sessions in the $R$ band and all 48 analysed
sessions in the $I$ band are classified as NV.

\item No statistically significant IDV is detected in
PG~1553+113. All 70 analysed sessions in the $R$ band and
all 66 analysed sessions in the $I$ band are classified as NV.

\item Robustness checks based on alternative comparison-star
pairings and on a simple Monte Carlo null test indicate that the
main IDV result is not driven by the adopted star pairing or by
the photometric noise level alone.

\item The distributions of the normalized F-test statistic remain
below the adopted variability threshold for 1ES~1959+650 and
PG~1553+113, whereas H~1426+428 shows a significant high-ratio
tail, confirming its higher incidence of optical IDV.

\end{itemize}

These results indicate that rapid optical synchrotron variability
in TeV HBLs is intermittent and strongly dependent on the activity
state of the source. The absence of robust IDV in 1ES~1959+650
and PG~1553+113, despite the extensive monitoring coverage,
shows that large long-term optical variability does not necessarily
imply frequent hour-scale variability. Both detections and
non-detections therefore provide useful constraints on the duty
cycle of rapid optical synchrotron variability in TeV blazar jets.

The internally consistent nature of the dataset, the use of two
independent comparison-star pairs, and the adopted classification
criteria provide conservative constraints on the duration-weighted
duty cycle of high-confidence optical IDV in these sources. In a
high-energy context, such optical constraints are relevant for
planning and interpreting multiwavelength campaigns of TeV HBLs,
especially when combined with simultaneous X-ray and gamma-ray
observations.

More generally, the present work illustrates the role that robotic
small- and medium-aperture optical telescopes can play in the
time-domain study of TeV blazars. For long-term monitoring and
duty-cycle studies, the availability of observing time, cadence,
repeatability, and homogeneous observing conditions can be as
important as aperture. The Asiago Schmidt programme therefore
provides a useful optical monitoring baseline for these and other
high-energy sources, and its continuation will extend the temporal
coverage, benefit from increasing automation of the reduction and
variability-analysis workflow, and support future integration into
broader multiwavelength and, where appropriate, multimessenger
collaborations.

\section*{Acknowledgements}
This work is based on observations collected at the Asiago Schmidt 67/92 telescope (Asiago Mount Ekar, Italy), operated by INAF -- Osservatorio Astronomico di Padova.\\
This research made use of the Starlink Software Collection
\citep{Currie2014}. The Starlink software is currently supported
by the East Asian Observatory.

\section*{Funding}

This research did not receive any specific grant from funding
agencies in the public, commercial, or not-for-profit sectors.

\section*{Declaration of competing interest}

The authors declare that they have no known competing financial
interests or personal relationships that could have appeared to
influence the work reported in this paper.

\bibliographystyle{elsarticle-num-names}
\bibliography{references}

@ARTICLE{Urry1995,
       author = {{Urry}, C. Megan and {Padovani}, Paolo},
        title = "{Unified Schemes for Radio-Loud Active Galactic Nuclei}",
      journal = {\pasp},
         year = 1995,
        month = sep,
       volume = {107},
        pages = {803},
          doi = {10.1086/133630},
archivePrefix = {arXiv},
       eprint = {astro-ph/9506063},
 primaryClass = {astro-ph},
       adsurl = {https://ui.adsabs.harvard.edu/abs/1995PASP..107..803U}
}

@ARTICLE{Fossati1998,
       author = {{Fossati}, G. and {Maraschi}, L. and {Celotti}, A. and {Comastri}, A. and {Ghisellini}, G.},
        title = "{A unifying view of the spectral energy distributions of blazars}",
      journal = {\mnras},
         year = 1998,
        month = sep,
       volume = {299},
       number = {2},
        pages = {433-448},
          doi = {10.1046/j.1365-8711.1998.01828.x},
archivePrefix = {arXiv},
       eprint = {astro-ph/9804103},
 primaryClass = {astro-ph},
       adsurl = {https://ui.adsabs.harvard.edu/abs/1998MNRAS.299..433F}
}

@ARTICLE{Ulrich1997,
       author = {{Ulrich}, Marie-Helene and {Maraschi}, Laura and {Urry}, C. Megan},
        title = "{Variability of Active Galactic Nuclei}",
      journal = {\araa},
         year = 1997,
        month = jan,
       volume = {35},
        pages = {445-502},
          doi = {10.1146/annurev.astro.35.1.445},
       adsurl = {https://ui.adsabs.harvard.edu/abs/1997ARA&A..35..445U}
}

@ARTICLE{Ghisellini2010,
       author = {{Ghisellini}, G. and {Tavecchio}, F. and {Foschini}, L. and {Ghirlanda}, G. and {Maraschi}, L. and {Celotti}, A.},
        title = "{General physical properties of bright Fermi blazars}",
      journal = {\mnras},
         year = 2010,
        month = feb,
       volume = {402},
       number = {1},
        pages = {497-518},
          doi = {10.1111/j.1365-2966.2009.15898.x},
archivePrefix = {arXiv},
       eprint = {0909.0932},
 primaryClass = {astro-ph.CO},
       adsurl = {https://ui.adsabs.harvard.edu/abs/2010MNRAS.402..497G}
}

@ARTICLE{Wagner1995,
       author = {{Wagner}, S.~J. and {Witzel}, A.},
        title = "{Intraday Variability In Quasars and BL Lac Objects}",
      journal = {\araa},
         year = 1995,
        month = jan,
       volume = {33},
        pages = {163-198},
          doi = {10.1146/annurev.aa.33.090195.001115},
       adsurl = {https://ui.adsabs.harvard.edu/abs/1995ARA&A..33..163W}
}

@ARTICLE{Perlman1996,
       author = {{Perlman}, Eric S. and {Stocke}, John T. and {Schachter}, Jonathan F. and {Elvis}, Martin and {Ellingson}, Erica and {Urry}, C. Megan and {Potter}, Michael and {Impey}, Chris D. and {Kolchinsky}, Peter},
        title = "{The Einstein Slew Survey Sample of BL Lacertae Objects}",
      journal = {\apjs},
         year = 1996,
        month = jun,
       volume = {104},
        pages = {251},
          doi = {10.1086/192300},
       adsurl = {https://ui.adsabs.harvard.edu/abs/1996ApJS..104..251P}
}

@ARTICLE{Padovani1995,
       author = {{Padovani}, Paolo and {Giommi}, Paolo},
        title = "{The Connection between X-Ray-- and Radio-selected BL Lacertae Objects}",
      journal = {\apj},
         year = 1995,
        month = may,
       volume = {444},
        pages = {567},
          doi = {10.1086/175631},
archivePrefix = {arXiv},
       eprint = {astro-ph/9412073},
 primaryClass = {astro-ph},
       adsurl = {https://ui.adsabs.harvard.edu/abs/1995ApJ...444..567P}
}

@ARTICLE{Ackermann2015,
       author = {{Ackermann}, M. and {Ajello}, M. and {Albert}, A. and {Atwood}, W.~B. and {Baldini}, L. and {Ballet}, J. and {Barbiellini}, G. and {Bastieri}, D. and {Becerra Gonzalez}, J. and {Bellazzini}, R. and {Bissaldi}, E. and {Blandford}, R.~D. and {Bloom}, E.~D. and {Bonino}, R. and {Bottacini}, E. and {Bregeon}, J. and {Bruel}, P. and {Buehler}, R. and {Buson}, S. and {Caliandro}, G.~A. and {Cameron}, R.~A. and {Caputo}, R. and {Caragiulo}, M. and {Caraveo}, P.~A. and {Cavazzuti}, E. and {Cecchi}, C. and {Chekhtman}, A. and {Chiang}, J. and {Chiaro}, G. and {Ciprini}, S. and {Cohen-Tanugi}, J. and {Conrad}, J. and {Cutini}, S. and {D'Ammando}, F. and {de Angelis}, A. and {de Palma}, F. and {Desiante}, R. and {Di Venere}, L. and {Dom{\'\i}nguez}, A. and {Drell}, P.~S. and {Favuzzi}, C. and {Fegan}, S.~J. and {Ferrara}, E.~C. and {Focke}, W.~B. and {Fuhrmann}, L. and {Fukazawa}, Y. and {Fusco}, P. and {Gargano}, F. and {Gasparrini}, D. and {Giglietto}, N. and {Giommi}, P. and {Giordano}, F. and {Giroletti}, M. and {Godfrey}, G. and {Green}, D. and {Grenier}, I.~A. and {Grove}, J.~E. and {Guiriec}, S. and {Harding}, A.~K. and {Hays}, E. and {Hewitt}, J.~W. and {Hill}, A.~B. and {Horan}, D. and {Jogler}, T. and {J{\'o}hannesson}, G. and {Johnson}, A.~S. and {Kamae}, T. and {Kuss}, M. and {Larsson}, S. and {Latronico}, L. and {Li}, J. and {Li}, L. and {Longo}, F. and {Loparco}, F. and {Lott}, B. and {Lovellette}, M.~N. and {Lubrano}, P. and {Magill}, J. and {Maldera}, S. and {Manfreda}, A. and {Max-Moerbeck}, W. and {Mayer}, M. and {Mazziotta}, M.~N. and {McEnery}, J.~E. and {Michelson}, P.~F. and {Mizuno}, T. and {Monzani}, M.~E. and {Morselli}, A. and {Moskalenko}, I.~V. and {Murgia}, S. and {Nuss}, E. and {Ohno}, M. and {Ohsugi}, T. and {Ojha}, R. and {Omodei}, N. and {Orlando}, E. and {Ormes}, J.~F. and {Paneque}, D. and {Pearson}, T.~J. and {Perkins}, J.~S. and {Perri}, M. and {Pesce-Rollins}, M. and {Petrosian}, V. and {Piron}, F. and {Pivato}, G. and {Porter}, T.~A. and {Rain{\`o}}, S. and {Rando}, R. and {Razzano}, M. and {Readhead}, A. and {Reimer}, A. and {Reimer}, O. and {Schulz}, A. and {Sgr{\`o}}, C. and {Siskind}, E.~J. and {Spada}, F. and {Spandre}, G. and {Spinelli}, P. and {Suson}, D.~J. and {Takahashi}, H. and {Thayer}, J.~B. and {Thompson}, D.~J. and {Tibaldo}, L. and {Torres}, D.~F. and {Tosti}, G. and {Troja}, E. and {Uchiyama}, Y. and {Vianello}, G. and {Wood}, K.~S. and {Wood}, M. and {Zimmer}, S. and {Berdyugin}, A. and {Corbet}, R.~H.~D. and {Hovatta}, T. and {Lindfors}, E. and {Nilsson}, K. and {Reinthal}, R. and {Sillanp{\"a}{\"a}}, A. and {Stamerra}, A. and {Takalo}, L.~O. and {Valtonen}, M.~J.},
        title = "{Multiwavelength Evidence for Quasi-periodic Modulation in the Gamma-Ray Blazar PG 1553+113}",
      journal = {\apjl},
         year = 2015,
        month = nov,
       volume = {813},
       number = {2},
          eid = {L41},
        pages = {L41},
          doi = {10.1088/2041-8205/813/2/L41},
archivePrefix = {arXiv},
       eprint = {1509.02063},
 primaryClass = {astro-ph.HE},
       adsurl = {https://ui.adsabs.harvard.edu/abs/2015ApJ...813L..41A}
}

@ARTICLE{Krawczynski2004,
       author = {{Krawczynski}, H. and {Hughes}, S.~B. and {Horan}, D. and {Aharonian}, F. and {Aller}, M.~F. and {Aller}, H. and {Boltwood}, P. and {Buckley}, J. and {Coppi}, P. and {Fossati}, G. and {G{\"o}tting}, N. and {Holder}, J. and {Horns}, D. and {Kurtanidze}, O.~M. and {Marscher}, A.~P. and {Nikolashvili}, M. and {Remillard}, R.~A. and {Sadun}, A. and {Schr{\"o}der}, M.},
        title = "{Multiwavelength Observations of Strong Flares from the TeV Blazar 1ES 1959+650}",
      journal = {\apj},
         year = 2004,
        month = jan,
       volume = {601},
       number = {1},
        pages = {151-164},
          doi = {10.1086/380393},
archivePrefix = {arXiv},
       eprint = {astro-ph/0310158},
 primaryClass = {astro-ph},
       adsurl = {https://ui.adsabs.harvard.edu/abs/2004ApJ...601..151K}
}

@ARTICLE{Costamante2001,
       author = {{Costamante}, L. and {Ghisellini}, G. and {Giommi}, P. and {Tagliaferri}, G. and {Celotti}, A. and {Chiaberge}, M. and {Fossati}, G. and {Maraschi}, L. and {Tavecchio}, F. and {Treves}, A. and {Wolter}, A.},
        title = "{Extreme synchrotron BL Lac objects. Stretching the blazar sequence}",
      journal = {\aap},
         year = 2001,
        month = may,
       volume = {371},
        pages = {512-526},
          doi = {10.1051/0004-6361:20010412},
archivePrefix = {arXiv},
       eprint = {astro-ph/0103343},
 primaryClass = {astro-ph},
       adsurl = {https://ui.adsabs.harvard.edu/abs/2001A&A...371..512C}
}

@ARTICLE{Danforth2010,
       author = {{Danforth}, Charles W. and {Keeney}, Brian A. and {Stocke}, John T. and {Shull}, J. Michael and {Yao}, Yangsen},
        title = "{Hubble/COS Observations of the Ly{\ensuremath{\alpha}} Forest Toward the BL Lac Object 1ES 1553+113}",
      journal = {\apj},
         year = 2010,
        month = sep,
       volume = {720},
       number = {1},
        pages = {976-986},
          doi = {10.1088/0004-637X/720/1/976},
archivePrefix = {arXiv},
       eprint = {1005.2191},
 primaryClass = {astro-ph.CO},
       adsurl = {https://ui.adsabs.harvard.edu/abs/2010ApJ...720..976D}
}

@ARTICLE{Donato2001,
       author = {{Donato}, D. and {Ghisellini}, G. and {Tagliaferri}, G. and {Fossati}, G.},
        title = "{Hard X-ray properties of blazars}",
      journal = {\aap},
         year = 2001,
        month = sep,
       volume = {375},
        pages = {739-751},
          doi = {10.1051/0004-6361:20010675},
archivePrefix = {arXiv},
       eprint = {astro-ph/0105203},
 primaryClass = {astro-ph},
       adsurl = {https://ui.adsabs.harvard.edu/abs/2001A&A...375..739D}
}

@ARTICLE{Su2024,
       author = {{Su}, Zi-An and {Yang}, Wen-Xin and {Zeng}, Xiang-Tao and {Ou}, Le-Jian and {Li}, Ze-Lin and {Yang}, Jiang-He and {Fan}, Jun-Hui},
        title = "{The Optical Variability Properties of TeV Blazars}",
      journal = {Research in Astronomy and Astrophysics},
         year = 2024,
        month = sep,
       volume = {24},
       number = {9},
          eid = {095005},
        pages = {095005},
          doi = {10.1088/1674-4527/ad6db4},
       adsurl = {https://ui.adsabs.harvard.edu/abs/2024RAA....24i5005S}
}

@ARTICLE{Carini1992,
       author = {{Carini}, M.~T. and {Miller}, H.~R. and {Noble}, J.~C. and {Goodrich}, B.~D.},
        title = "{The Timescales of the Optical Variability of Blazars. III. OJ 287 and BL Lacertae}",
      journal = {\aj},
         year = 1992,
        month = jul,
       volume = {104},
        pages = {15},
          doi = {10.1086/116217},
       adsurl = {https://ui.adsabs.harvard.edu/abs/1992AJ....104...15C}
}

@ARTICLE{Meng2017,
       author = {{Meng}, Nankun and {Wu}, Jianghua and {Webb}, James R. and {Zhang}, Xiaoyuan and {Dai}, Yan},
        title = "{Intraday optical variability of BL Lacertae}",
      journal = {\mnras},
         year = 2017,
        month = aug,
       volume = {469},
       number = {3},
        pages = {3588-3596},
          doi = {10.1093/mnras/stx1055},
archivePrefix = {arXiv},
       eprint = {1706.07183},
 primaryClass = {astro-ph.GA},
       adsurl = {https://ui.adsabs.harvard.edu/abs/2017MNRAS.469.3588M}
}

@ARTICLE{Fang2022,
       author = {{Fang}, Yue and {Zhang}, Yan and {Chen}, Qihang and {Wu}, Jianghua},
        title = "{Intraday Optical Multiband Observation of BL Lacertae}",
      journal = {\apj},
         year = 2022,
        month = feb,
       volume = {926},
       number = {1},
          eid = {91},
        pages = {91},
          doi = {10.3847/1538-4357/ac4490},
       adsurl = {https://ui.adsabs.harvard.edu/abs/2022ApJ...926...91F}
}

@ARTICLE{Tavani2018,
       author = {{Tavani}, M. and {Cavaliere}, A. and {Munar-Adrover}, Pere and {Argan}, A.},
        title = "{The Blazar PG 1553+113 as a Binary System of Supermassive Black Holes}",
      journal = {\apj},
         year = 2018,
        month = feb,
       volume = {854},
       number = {1},
          eid = {11},
        pages = {11},
          doi = {10.3847/1538-4357/aaa3f4},
archivePrefix = {arXiv},
       eprint = {1801.03335},
 primaryClass = {astro-ph.HE},
       adsurl = {https://ui.adsabs.harvard.edu/abs/2018ApJ...854...11T}
}

@ARTICLE{deDiego2010,
       author = {{de Diego}, Jos{\'e} A.},
        title = "{Testing Tests on Active Galactic Nucleus Microvariability}",
      journal = {\aj},
         year = 2010,
        month = mar,
       volume = {139},
       number = {3},
        pages = {1269-1282},
          doi = {10.1088/0004-6256/139/3/1269},
archivePrefix = {arXiv},
       eprint = {1001.2543},
 primaryClass = {astro-ph.CO},
       adsurl = {https://ui.adsabs.harvard.edu/abs/2010AJ....139.1269D}
}

@ARTICLE{Aharonian2002,
       author = {{Aharonian}, F. and {Akhperjanian}, A. and {Barrio}, J. and {Beilicke}, M. and {Bernl{\"o}hr}, K. and {B{\"o}rst}, H. and {Bojahr}, H. and {Bolz}, O. and {Contreras}, J. and {Cornils}, R. and {Cortina}, J. and {Denninghoff}, S. and {Fonseca}, V. and {Girma}, M. and {Gonzalez}, J. and {G{\"o}tting}, N. and {Heinzelmann}, G. and {Hermann}, G. and {Heusler}, A. and {Hofmann}, W. and {Horns}, D. and {Jung}, I. and {Kankanyan}, R. and {Kestel}, M. and {Kettler}, J. and {Kohnle}, A. and {Konopelko}, A. and {Kornmeyer}, H. and {Kranich}, D. and {Krawczynski}, H. and {Lampeitl}, H. and {Lopez}, M. and {Lorenz}, E. and {Lucarelli}, F. and {Magnussen}, N. and {Mang}, O. and {Meyer}, H. and {Mirzoyan}, R. and {Moralejo}, A. and {Ona}, E. and {Padilla}, L. and {Panter}, M. and {Plaga}, R. and {Plyasheshnikov}, A. and {P{\"u}hlhofer}, G. and {Rauterberg}, G. and {R{\"o}hring}, A. and {Rhode}, W. and {Robrade}, J. and {Rowell}, G. and {Sahakian}, V. and {Samorski}, M. and {Schilling}, M. and {Schr{\"o}der}, F. and {Sevilla}, I. and {Siems}, M. and {Stamm}, W. and {Tluczykont}, M. and {V{\"o}lk}, H.~J. and {Wiedner}, C.~A. and {Wittek}, W.},
        title = "{TeV gamma rays from the blazar H 1426+428 and the diffuse extragalactic background radiation}",
      journal = {\aap},
         year = 2002,
        month = mar,
       volume = {384},
        pages = {L23-L26},
          doi = {10.1051/0004-6361:20020206},
       adsurl = {https://ui.adsabs.harvard.edu/abs/2002A&A...384L..23A}
}

@ARTICLE{Aharonian2006,
       author = {{Aharonian}, F. and {Akhperjanian}, A.~G. and {Bazer-Bachi}, A.~R. and {Beilicke}, M. and {Benbow}, W. and {Berge}, D. and {Bernl{\"o}hr}, K. and {Boisson}, C. and {Bolz}, O. and {Borrel}, V. and {Braun}, I. and {Breitling}, F. and {Brown}, A.~M. and {B{\"u}hler}, R. and {Carrigan}, S. and {Chadwick}, P.~M. and {Chounet}, L.-M. and {Cornils}, R. and {Costamante}, L. and {Degrange}, B. and {Dickinson}, H.~J. and {Djannati-Ata{\"\i}}, A. and {O'C. Drury}, L. and {Dubus}, G. and {Egberts}, K. and {Emmanoulopoulos}, D. and {Espigat}, P. and {Feinstein}, F. and {Fontaine}, G. and {Funk}, S. and {Gallant}, Y.~A. and {Giebels}, B. and {Glicenstein}, J.~F. and {Goret}, P. and {Hadjichristidis}, C. and {Hauser}, D. and {Hauser}, M. and {Heinzelmann}, G. and {Henri}, G. and {Hermann}, G. and {Hinton}, J.~A. and {Hofmann}, W. and {Holleran}, M. and {Horns}, D. and {Jacholkowska}, A. and {de Jager}, O.~C. and {Kh{\'e}lifi}, B. and {Komin}, Nu. and {Konopelko}, A. and {Latham}, I.~J. and {Le Gallou}, R. and {Lemi{\`e}re}, A. and {Lemoine-Goumard}, M. and {Lohse}, T. and {Martin}, J.~M. and {Martineau-Huynh}, O. and {Marcowith}, A. and {Masterson}, C. and {McComb}, T.~J.~L. and {de Naurois}, M. and {Nedbal}, D. and {Nolan}, S.~J. and {Noutsos}, A. and {Orford}, K.~J. and {Osborne}, J.~L. and {Ouchrif}, M. and {Panter}, M. and {Pelletier}, G. and {Pita}, S. and {P{\"u}hlhofer}, G. and {Punch}, M. and {Raubenheimer}, B.~C. and {Raue}, M. and {Rayner}, S.~M. and {Reimer}, A. and {Reimer}, O. and {Ripken}, J. and {Rob}, L. and {Rolland}, L. and {Rowell}, G. and {Sahakian}, V. and {Saug{\'e}}, L. and {Schlenker}, S. and {Schlickeiser}, R. and {Schuster}, C. and {Schwanke}, U. and {Siewert}, M. and {Sol}, H. and {Spangler}, D. and {Steenkamp}, R. and {Stegmann}, C. and {Superina}, G. and {Tavernet}, J.-P. and {Terrier}, R. and {Th{\'e}oret}, C.~G. and {Tluczykont}, M. and {van Eldik}, C. and {Vasileiadis}, G. and {Venter}, C. and {Vincent}, P. and {V{\"o}lk}, H.~J. and {Wagner}, S.~J. and {Ward}, M.},
        title = "{Evidence for VHE {\ensuremath{\gamma}}-ray emission from the distant BL Lac PG 1553+113}",
      journal = {\aap},
         year = 2006,
        month = mar,
       volume = {448},
       number = {2},
        pages = {L19-L23},
          doi = {10.1051/0004-6361:200600010},
archivePrefix = {arXiv},
       eprint = {astro-ph/0601545},
 primaryClass = {astro-ph},
       adsurl = {https://ui.adsabs.harvard.edu/abs/2006A&A...448L..19A}
}

@ARTICLE{Racine1970,
       author = {{Racine}, Ren{\'e}},
        title = "{Photometry of BL Lacertae with a Time Resolution of 15 Seconds}",
      journal = {\apjl},
         year = 1970,
        month = feb,
       volume = {159},
        pages = {L99},
          doi = {10.1086/180486},
       adsurl = {https://ui.adsabs.harvard.edu/abs/1970ApJ...159L..99R}
}

@ARTICLE{Abdo2009,
       author = {{Abdo}, A.~A. and {Ackermann}, M. and {Ajello}, M. and {Atwood}, W.~B. and {Axelsson}, M. and {Baldini}, L. and {Ballet}, J. and {Barbiellini}, G. and {Baring}, M.~G. and {Bastieri}, D. and {Bechtol}, K. and {Bellazzini}, R. and {Berenji}, B. and {Bloom}, E.~D. and {Bonamente}, E. and {Borgland}, A.~W. and {Bregeon}, J. and {Brez}, A. and {Brigida}, M. and {Bruel}, P. and {Burnett}, T.~H. and {Caliandro}, G.~A. and {Cameron}, R.~A. and {Caraveo}, P.~A. and {Casandjian}, J.~M. and {Cavazzuti}, E. and {Cecchi}, C. and {{\c{C}}elik}, {\"O}. and {Chekhtman}, A. and {Cheung}, C.~C. and {Chiang}, J. and {Ciprini}, S. and {Claus}, R. and {Cohen-Tanugi}, J. and {Cominsky}, L.~R. and {Conrad}, J. and {Cutini}, S. and {de Angelis}, A. and {de Palma}, F. and {Di Bernardo}, G. and {Silva}, E. do Couto e. and {Drell}, P.~S. and {Drlica-Wagner}, A. and {Dubois}, R. and {Dumora}, D. and {Farnier}, C. and {Favuzzi}, C. and {Fegan}, S.~J. and {Finke}, J. and {Focke}, W.~B. and {Fortin}, P. and {Foschini}, L. and {Frailis}, M. and {Fukazawa}, Y. and {Funk}, S. and {Fusco}, P. and {Gargano}, F. and {Gasparrini}, D. and {Gehrels}, N. and {Germani}, S. and {Giavitto}, G. and {Giebels}, B. and {Giglietto}, N. and {Giommi}, P. and {Giordano}, F. and {Glanzman}, T. and {Godfrey}, G. and {Grenier}, I.~A. and {Grondin}, M.-H. and {Grove}, J.~E. and {Guillemot}, L. and {Guiriec}, S. and {Hanabata}, Y. and {Hayashida}, M. and {Hays}, E. and {Horan}, D. and {Hughes}, R.~E. and {Jackson}, M.~S. and {J{\'o}hannesson}, G. and {Johnson}, A.~S. and {Johnson}, R.~P. and {Johnson}, W.~N. and {Kamae}, T. and {Katagiri}, H. and {Kataoka}, J. and {Kawai}, N. and {Kerr}, M. and {Kn{\"o}dlseder}, J. and {Kocian}, M.~L. and {Kuss}, M. and {Lande}, J. and {Latronico}, L. and {Lemoine-Goumard}, M. and {Longo}, F. and {Loparco}, F. and {Lott}, B. and {Lovellette}, M.~N. and {Lubrano}, P. and {Madejski}, G.~M. and {Makeev}, A. and {Mazziotta}, M.~N. and {McConville}, W. and {McEnery}, J.~E. and {Meurer}, C. and {Michelson}, P.~F. and {Mitthumsiri}, W. and {Mizuno}, T. and {Moiseev}, A.~A. and {Monte}, C. and {Monzani}, M.~E. and {Morselli}, A. and {Moskalenko}, I.~V. and {Murgia}, S. and {Nolan}, P.~L. and {Norris}, J.~P. and {Nuss}, E. and {Ohsugi}, T. and {Omodei}, N. and {Orlando}, E. and {Ormes}, J.~F. and {Ozaki}, M. and {Paneque}, D. and {Panetta}, J.~H. and {Parent}, D. and {Pelassa}, V. and {Pepe}, M. and {Pesce-Rollins}, M. and {Piron}, F. and {Porter}, T.~A. and {Rain{\`o}}, S. and {Rando}, R. and {Razzano}, M. and {Reimer}, A. and {Reimer}, O. and {Reposeur}, T. and {Reyes}, L.~C. and {Ritz}, S. and {Rochester}, L.~S. and {Rodriguez}, A.~Y. and {Roth}, M. and {Ryde}, F. and {Sadrozinski}, H.~F.-W. and {Sanchez}, D. and {Sander}, A. and {Saz Parkinson}, P.~M. and {Scargle}, J.~D. and {Schalk}, T.~L. and {Sellerholm}, A. and {Sgr{\`o}}, C. and {Shaw}, M.~S. and {Siskind}, E.~J. and {Smith}, D.~A. and {Smith}, P.~D. and {Spandre}, G. and {Spinelli}, P. and {Strickman}, M.~S. and {Suson}, D.~J. and {Tajima}, H. and {Takahashi}, H. and {Takahashi}, T. and {Tanaka}, T. and {Tanaka}, Y. and {Thayer}, J.~B. and {Thayer}, J.~G. and {Thompson}, D.~J. and {Tibaldo}, L. and {Torres}, D.~F. and {Tosti}, G. and {Tramacere}, A. and {Uchiyama}, Y. and {Usher}, T.~L. and {Vasileiou}, V. and {Vilchez}, N. and {Vitale}, V. and {Waite}, A.~P. and {Wang}, P. and {Winer}, B.~L. and {Wood}, K.~S. and {Ylinen}, T. and {Ziegler}, M.},
        title = "{Fermi Observations of TeV-Selected Active Galactic Nuclei}",
      journal = {\apj},
         year = 2009,
        month = dec,
       volume = {707},
       number = {2},
        pages = {1310-1333},
          doi = {10.1088/0004-637X/707/2/1310},
archivePrefix = {arXiv},
       eprint = {0910.4881},
 primaryClass = {astro-ph.HE},
       adsurl = {https://ui.adsabs.harvard.edu/abs/2009ApJ...707.1310A}
}

@ARTICLE{Horan2002,
       author = {{Horan}, D. and {Badran}, H.~M. and {Bond}, I.~H. and {Bradbury}, S.~M. and {Buckley}, J.~H. and {Carson}, M.~J. and {Carter-Lewis}, D.~A. and {Catanese}, M. and {Cui}, W. and {Dunlea}, S. and {Das}, D. and {de la Calle Perez}, I. and {D'Vali}, M. and {Fegan}, D.~J. and {Fegan}, S.~J. and {Finley}, J.~P. and {Gaidos}, J.~A. and {Gibbs}, K. and {Gillanders}, G.~H. and {Hall}, T.~A. and {Hillas}, A.~M. and {Holder}, J. and {Jordan}, M. and {Kertzman}, M. and {Kieda}, D. and {Kildea}, J. and {Knapp}, J. and {Kosack}, K. and {Krennrich}, F. and {Lang}, M.~J. and {LeBohec}, S. and {Lessard}, R. and {Lloyd-Evans}, J. and {McKernan}, B. and {Moriarty}, P. and {Muller}, D. and {Ong}, R. and {Pallassini}, R. and {Petry}, D. and {Quinn}, J. and {Reay}, N.~W. and {Reynolds}, P.~T. and {Rose}, H.~J. and {Sembroski}, G.~H. and {Sidwell}, R. and {Stanton}, N. and {Swordy}, S.~P. and {Vassiliev}, V.~V. and {Wakely}, S.~P. and {Weekes}, T.~C.},
        title = "{Detection of the BL Lacertae Object H1426+428 at TeV Gamma-Ray Energies}",
      journal = {\apj},
         year = 2002,
        month = jun,
       volume = {571},
       number = {2},
        pages = {753-762},
          doi = {10.1086/340019},
archivePrefix = {arXiv},
       eprint = {astro-ph/0202185},
 primaryClass = {astro-ph},
       adsurl = {https://ui.adsabs.harvard.edu/abs/2002ApJ...571..753H}
}

@ARTICLE{Johnson2019,
       author = {{Johnson}, Sean D. and {Mulchaey}, John S. and {Chen}, Hsiao-Wen and {Wijers}, Nastasha A. and {Connor}, Thomas and {Muzahid}, Sowgat and {Schaye}, Joop and {Cen}, Renyue and {Carlsten}, Scott G. and {Charlton}, Jane and {Drout}, Maria R. and {Goulding}, Andy D. and {Hansen}, Terese T. and {Walth}, Gregory L.},
        title = "{The Physical Origins of the Identified and Still Missing Components of the Warm-Hot Intergalactic Medium: Insights from Deep Surveys in the Field of Blazar 1ES1553+113}",
      journal = {\apjl},
         year = 2019,
        month = oct,
       volume = {884},
       number = {2},
          eid = {L31},
        pages = {L31},
          doi = {10.3847/2041-8213/ab479a},
archivePrefix = {arXiv},
       eprint = {1909.11667},
 primaryClass = {astro-ph.GA},
       adsurl = {https://ui.adsabs.harvard.edu/abs/2019ApJ...884L..31J}
}

@ARTICLE{Prandini2010,
       author = {{Prandini}, E. and {Bonnoli}, G. and {Maraschi}, L. and {Mariotti}, M. and {Tavecchio}, F.},
        title = "{Constraining blazar distances with combined Fermi and TeV data: an empirical approach}",
      journal = {\mnras},
         year = 2010,
        month = jun,
       volume = {405},
       number = {1},
        pages = {L76-L80},
          doi = {10.1111/j.1745-3933.2010.00862.x},
archivePrefix = {arXiv},
       eprint = {1003.1674},
 primaryClass = {astro-ph.CO},
       adsurl = {https://ui.adsabs.harvard.edu/abs/2010MNRAS.405L..76P}
}

@ARTICLE{Aliu2013,
       author = {{Aliu}, E. and {Archambault}, S. and {Arlen}, T. and {Aune}, T. and {Beilicke}, M. and {Benbow}, W. and {Bird}, R. and {B{\"o}ttcher}, M. and {Bouvier}, A. and {Bugaev}, V. and {Byrum}, K. and {Cesarini}, A. and {Ciupik}, L. and {Collins-Hughes}, E. and {Connolly}, M.~P. and {Cui}, W. and {Dickherber}, R. and {Duke}, C. and {Dumm}, J. and {Errando}, M. and {Falcone}, A. and {Federici}, S. and {Feng}, Q. and {Finley}, J.~P. and {Finnegan}, G. and {Fortson}, L. and {Furniss}, A. and {Galante}, N. and {Gall}, D. and {Gillanders}, G.~H. and {Griffin}, S. and {Grube}, J. and {Gyuk}, G. and {Hanna}, D. and {Holder}, J. and {Hughes}, G. and {Humensky}, T.~B. and {Kaaret}, P. and {Kertzman}, M. and {Khassen}, Y. and {Kieda}, D. and {Krawczynski}, H. and {Krennrich}, F. and {Lang}, M.~J. and {Madhavan}, A.~S. and {Maier}, G. and {Majumdar}, P. and {McArthur}, S. and {McCann}, A. and {Moriarty}, P. and {Mukherjee}, R. and {Nelson}, T. and {O'Faol{\'a}in de Bhr{\'o}ithe}, A. and {Ong}, R.~A. and {Orr}, M. and {Otte}, A.~N. and {Park}, N. and {Perkins}, J.~S. and {Pichel}, A. and {Pohl}, M. and {Popkow}, A. and {Prokoph}, H. and {Quinn}, J. and {Ragan}, K. and {Reyes}, L.~C. and {Reynolds}, P.~T. and {Roache}, E. and {Saxon}, D.~B. and {Schroedter}, M. and {Sembroski}, G.~H. and {Skole}, C. and {Smith}, A.~W. and {Staszak}, D. and {Telezhinsky}, I. and {Theiling}, M. and {Tyler}, J. and {Varlotta}, A. and {Vassiliev}, V.~V. and {Wakely}, S.~P. and {Weekes}, T.~C. and {Weinstein}, A. and {Welsing}, R. and {Williams}, D.~A. and {Zitzer}, B.},
        title = "{Multiwavelength Observations and Modeling of 1ES 1959+650 in a Low Flux State}",
      journal = {\apj},
         year = 2013,
        month = sep,
       volume = {775},
       number = {1},
          eid = {3},
        pages = {3},
          doi = {10.1088/0004-637X/775/1/3},
archivePrefix = {arXiv},
       eprint = {1307.6772},
 primaryClass = {astro-ph.HE},
       adsurl = {https://ui.adsabs.harvard.edu/abs/2013ApJ...775....3A}
}

@ARTICLE{Boettcher2005,
       author = {{B{\"o}ttcher}, Markus},
        title = "{A Hadronic Synchrotron Mirror Model for the ``Orphan'' TeV Flare in 1ES 1959+650}",
      journal = {\apj},
         year = 2005,
        month = mar,
       volume = {621},
       number = {1},
        pages = {176-180},
          doi = {10.1086/427430},
archivePrefix = {arXiv},
       eprint = {astro-ph/0411248},
 primaryClass = {astro-ph},
       adsurl = {https://ui.adsabs.harvard.edu/abs/2005ApJ...621..176B}
}

@ARTICLE{Remillard1989,
       author = {{Remillard}, R.~A. and {Tuohy}, I.~R. and {Brissenden}, R.~J.~V. and {Buckley}, D.~A.~H. and {Schwartz}, D.~A. and {Feigelson}, E.~D. and {Tapia}, S.},
        title = "{Two X-Ray--selected BL Lacertae Objects Observed with the HEAO 1 Scanning Modulation Collimator}",
      journal = {\apj},
         year = 1989,
        month = oct,
       volume = {345},
        pages = {140},
          doi = {10.1086/167888},
       adsurl = {https://ui.adsabs.harvard.edu/abs/1989ApJ...345..140R}
}

@ARTICLE{Kapanadze2016,
       author = {{Kapanadze}, B. and {Romano}, P. and {Vercellone}, S. and {Kapanadze}, S. and {Mdzinarishvili}, T. and {Kharshiladze}, G.},
        title = "{The long-term Swift observations of the high-energy peaked BL Lacertae source 1ES 1959+650}",
      journal = {\mnras},
         year = 2016,
        month = mar,
       volume = {457},
       number = {1},
        pages = {704-722},
          doi = {10.1093/mnras/stv3004},
       adsurl = {https://ui.adsabs.harvard.edu/abs/2016MNRAS.457..704K}
}

@ARTICLE{Pandey2020,
       author = {{Pandey}, Ashwani and {Gupta}, Alok C. and {Kurtanidze}, Sofia O. and {Wiita}, Paul J. and {Damljanovic}, G. and {Bachev}, R. and {Zhang}, Jin and {Kurtanidze}, O.~M. and {Darriba}, A. and {Chigladze}, R.~A. and {Latev}, G. and {Nikolashvili}, M.~G. and {Peneva}, S. and {Semkov}, E. and {Strigachev}, A. and {Tiwari}, S.~N. and {Vince}, O.},
        title = "{Optical Variability of the TeV Blazar 1ES 0806+524 on Diverse Timescales}",
      journal = {\apj},
         year = 2020,
        month = feb,
       volume = {890},
       number = {1},
          eid = {72},
        pages = {72},
          doi = {10.3847/1538-4357/ab698e},
archivePrefix = {arXiv},
       eprint = {2001.02398},
 primaryClass = {astro-ph.HE},
       adsurl = {https://ui.adsabs.harvard.edu/abs/2020ApJ...890...72P}
}

@ARTICLE{Massaro2011,
       author = {{Massaro}, F. and {Paggi}, A. and {Cavaliere}, A.},
        title = "{X-Ray and TeV Emissions from High-frequency-peaked BL Lac Objects}",
      journal = {\apjl},
         year = 2011,
        month = dec,
       volume = {742},
       number = {2},
          eid = {L32},
        pages = {L32},
          doi = {10.1088/2041-8205/742/2/L32},
archivePrefix = {arXiv},
       eprint = {1203.1924},
 primaryClass = {astro-ph.HE},
       adsurl = {https://ui.adsabs.harvard.edu/abs/2011ApJ...742L..32M}
}

@ARTICLE{Djannati2002,
       author = {{Djannati-Ata{\"\i}}, A. and {Khelifi}, B. and {Vorobiov}, S. and {Bazer-Bachi}, R. and {Chounet}, L.~M. and {Debiais}, G. and {Degrange}, B. and {Espigat}, P. and {Fabre}, B. and {Fontaine}, G. and {Goret}, P. and {Gouiffes}, C. and {Masterson}, C. and {Piron}, F. and {Punch}, M. and {Rivoal}, M. and {Rob}, L. and {Tavernet}, J.-P.},
        title = "{Detection of the BL Lac object 1ES 1426+428 in the Very High Energy gamma-ray band by the CAT Telescope from 1998-2000}",
      journal = {\aap},
         year = 2002,
        month = aug,
       volume = {391},
        pages = {L25-L28},
          doi = {10.1051/0004-6361:20021034},
archivePrefix = {arXiv},
       eprint = {astro-ph/0207618},
 primaryClass = {astro-ph},
       adsurl = {https://ui.adsabs.harvard.edu/abs/2002A&A...391L..25D}
}

@ARTICLE{Mazin2007,
       author = {{Mazin}, Daniel and {Goebel}, Florian},
        title = "{Break in the Very High Energy Spectrum of PG 1553+113: New Upper Limit on Its Redshift?}",
      journal = {\apjl},
         year = 2007,
        month = jan,
       volume = {655},
       number = {1},
        pages = {L13-L16},
          doi = {10.1086/511751},
archivePrefix = {arXiv},
       eprint = {astro-ph/0611817},
 primaryClass = {astro-ph},
       adsurl = {https://ui.adsabs.harvard.edu/abs/2007ApJ...655L..13M}
}

@ARTICLE{Giebels2002,
       author = {{Giebels}, Berrie and {Bloom}, Elliott D. and {Focke}, Warren and {Godfrey}, Gary and {Madejski}, Greg and {Reilly}, Kaice T. and {Parkinson}, Pablo M. Saz and {Shabad}, Ganya and {Bandyopadhyay}, Reba M. and {Fritz}, Gilbert G. and {Hertz}, Paul and {Kowalski}, Michael P. and {Lovellette}, Michael N. and {Ray}, Paul S. and {Wolff}, Michael T. and {Wood}, Kent S. and {Yentis}, Daryl J. and {Scargle}, Jeffrey D.},
        title = "{Observation of X-Ray Variability in the BL Lacertae Object 1ES 1959+65}",
      journal = {\apj},
         year = 2002,
        month = jun,
       volume = {571},
       number = {2},
        pages = {763-770},
          doi = {10.1086/340065},
archivePrefix = {arXiv},
       eprint = {astro-ph/0203335},
 primaryClass = {astro-ph},
       adsurl = {https://ui.adsabs.harvard.edu/abs/2002ApJ...571..763G}
}

@ARTICLE{Tagliaferri2003,
       author = {{Tagliaferri}, G. and {Ravasio}, M. and {Ghisellini}, G. and {Tavecchio}, F. and {Giommi}, P. and {Massaro}, E. and {Nesci}, R. and {Tosti}, G.},
        title = "{Wide band X-ray and optical observations of the BL Lac object 1ES 1959+650 in high state}",
      journal = {\aap},
         year = 2003,
        month = dec,
       volume = {412},
        pages = {711-720},
          doi = {10.1051/0004-6361:20034051},
archivePrefix = {arXiv},
       eprint = {astro-ph/0309568},
 primaryClass = {astro-ph},
       adsurl = {https://ui.adsabs.harvard.edu/abs/2003A&A...412..711T}
}

@ARTICLE{Cellone2007,
       author = {{Cellone}, Sergio A. and {Romero}, Gustavo E. and {Combi}, Jorge A. and {Mart{\'\i}}, Josep},
        title = "{Extreme photopolarimetric behaviour of the blazar AO0235+164}",
      journal = {\mnras},
         year = 2007,
        month = oct,
       volume = {381},
       number = {1},
        pages = {L60-L64},
          doi = {10.1111/j.1745-3933.2007.00366.x},
archivePrefix = {arXiv},
       eprint = {0707.2599},
 primaryClass = {astro-ph},
       adsurl = {https://ui.adsabs.harvard.edu/abs/2007MNRAS.381L..60C}
}

@ARTICLE{Kapanadze2012,
       author = {{Kapanadze}, Bidzina Z. and {Janiashvili}, Eduard},
        title = "{Variable synchrotron emission from BL Lacertae objects. II. Optical and X-ray flares in HBL source 1ES 1959+650}",
      journal = {\apss},
         year = 2012,
        month = jun,
       volume = {339},
       number = {2},
        pages = {339-353},
          doi = {10.1007/s10509-012-0984-y},
       adsurl = {https://ui.adsabs.harvard.edu/abs/2012Ap&SS.339..339K}
}

@ARTICLE{Bachev2012,
       author = {{Bachev}, R. and {Semkov}, E. and {Strigachev}, A. and {Gupta}, Alok C. and {Gaur}, Haritma and {Mihov}, B. and {Boeva}, S. and {Slavcheva-Mihova}, L.},
        title = "{The nature of the intra-night optical variability in blazars}",
      journal = {\mnras},
         year = 2012,
        month = aug,
       volume = {424},
       number = {4},
        pages = {2625-2634},
          doi = {10.1111/j.1365-2966.2012.21310.x},
archivePrefix = {arXiv},
       eprint = {1207.5945},
 primaryClass = {astro-ph.HE},
       adsurl = {https://ui.adsabs.harvard.edu/abs/2012MNRAS.424.2625B}
}

@INPROCEEDINGS{Costamante2002,
       author = {{Costamante}, L. and {Ghisellini}, G. and {Celotti}, A. and {Wolter}, A.},
        title = "{The BeppoSAX View of Extreme BL Lacs}",
    booktitle = {Blazar Astrophysics with BeppoSAX and Other Observatories},
         year = 2002,
       editor = {{Giommi}, Paolo and {Massaro}, Enrico and {Palumbo}, Giorgio},
        month = jan,
        pages = {21},
          doi = {10.48550/arXiv.astro-ph/0206482},
archivePrefix = {arXiv},
       eprint = {astro-ph/0206482},
 primaryClass = {astro-ph},
       adsurl = {https://ui.adsabs.harvard.edu/abs/2002babs.conf...21C}
}

@ARTICLE{Hinton2009,
       author = {{Hinton}, Jim},
        title = "{Ground-based gamma-ray astronomy with Cherenkov telescopes}",
      journal = {New Journal of Physics},
         year = 2009,
        month = may,
       volume = {11},
       number = {5},
          eid = {055005},
        pages = {055005},
          doi = {10.1088/1367-2630/11/5/055005},
archivePrefix = {arXiv},
       eprint = {0803.1609},
 primaryClass = {astro-ph},
       adsurl = {https://ui.adsabs.harvard.edu/abs/2009NJPh...11e5005H}
}

@ARTICLE{Bonnoli2015,
       author = {{Bonnoli}, G. and {Tavecchio}, F. and {Ghisellini}, G. and {Sbarrato}, T.},
        title = "{An emerging population of BL Lacs with extreme properties: towards a class of EBL and cosmic magnetic field probes?}",
      journal = {\mnras},
         year = 2015,
        month = jul,
       volume = {451},
       number = {1},
        pages = {611-621},
          doi = {10.1093/mnras/stv953},
archivePrefix = {arXiv},
       eprint = {1501.01974},
 primaryClass = {astro-ph.HE},
       adsurl = {https://ui.adsabs.harvard.edu/abs/2015MNRAS.451..611B}
}

@ARTICLE{Zhang2017,
       author = {{Zhang}, You-Hong and {Li}, Jia-Chen},
        title = "{Optical variability of the high synchrotron energy peaked blazar 1ES 1959+650 on various time-scales}",
      journal = {\mnras},
         year = 2017,
        month = aug,
       volume = {469},
       number = {2},
        pages = {1682-1690},
          doi = {10.1093/mnras/stx942},
       adsurl = {https://ui.adsabs.harvard.edu/abs/2017MNRAS.469.1682Z}
}

@ARTICLE{Dhiman2023,
       author = {{Dhiman}, Vinit and {Gupta}, Alok C. and {Kurtanidze}, Sofia O. and {Eglitis}, I. and {Strigachev}, A. and {Damljanovic}, G. and {Wiita}, Paul J. and {Gu}, Minfeng and {Gaur}, Haritma and {Vince}, Oliver and {Bachev}, R. and {Bisen}, D.~P. and {Ibryamov}, S. and {Ivanidze}, R.~Z. and {Jovanovic}, Miljana D. and {Kurtanidze}, Omar M. and {Nikolashvili}, M.~G. and {Semkov}, E. and {Spassov}, B. and {Stojanovic}, M. and {Villarroel}, Beatriz and {Xu}, Haiguang and {Zhang}, Zhongli},
        title = "{Multiband optical variability of the TeV blazar PG 1553 + 113 in 2019}",
      journal = {\mnras},
         year = 2023,
        month = feb,
       volume = {519},
       number = {2},
        pages = {2796-2811},
          doi = {10.1093/mnras/stac3709},
archivePrefix = {arXiv},
       eprint = {2212.09016},
 primaryClass = {astro-ph.HE},
       adsurl = {https://ui.adsabs.harvard.edu/abs/2023MNRAS.519.2796D}
}

@ARTICLE{Izviekova2022,
  author  = {Izviekova, I.~O. and Ponomarenko, V.~A. and Pulatova, N.~G. and Vasylenko, V.~V. and Simon, A.~O.},
  title   = {Photometric Variability of BL Lacertae and 1ES 1426+428 Blazars in the Optical and Gamma Ranges},
  journal = {Kinematics and Physics of Celestial Bodies},
  year    = {2022},
  volume  = {38},
  number  = {6},
  pages   = {328--339},
  doi     = {10.3103/S0884591322060034}
}

@ARTICLE{Nilsson2007,
       author = {{Nilsson}, K. and {Pasanen}, M. and {Takalo}, L.~O. and {Lindfors}, E. and {Berdyugin}, A. and {Ciprini}, S. and {Pforr}, J.},
        title = "{Host galaxy subtraction of TeV candidate BL Lacertae objects}",
      journal = {\aap},
         year = 2007,
        month = nov,
       volume = {475},
       number = {1},
        pages = {199-207},
          doi = {10.1051/0004-6361:20077624},
archivePrefix = {arXiv},
       eprint = {0709.2533},
 primaryClass = {astro-ph},
       adsurl = {https://ui.adsabs.harvard.edu/abs/2007A&A...475..199N}
}

@ARTICLE{Pace2013,
       author = {{Pace}, Cameron J. and {Pearson}, III, Richard L. and {Moody}, J. Ward and {Joner}, Michael D. and {Little}, Bret},
        title = "{The Stability of BVRI Comparison Stars Near Selected TeV Blazars}",
      journal = {\pasp},
         year = 2013,
        month = apr,
       volume = {125},
       number = {926},
        pages = {344},
          doi = {10.1086/670381},
archivePrefix = {arXiv},
       eprint = {1303.1380},
 primaryClass = {astro-ph.SR},
       adsurl = {https://ui.adsabs.harvard.edu/abs/2013PASP..125..344P}
}

@ARTICLE{Smith1991,
       author = {{Smith}, Paul S. and {Jannuzi}, Buell T. and {Elston}, Richard},
        title = "{UBVRI Photometry of Stars in the Fields of X-Ray Selected BL Lacertae Objects}",
      journal = {\apjs},
         year = 1991,
        month = sep,
       volume = {77},
        pages = {67},
          doi = {10.1086/191598},
       adsurl = {https://ui.adsabs.harvard.edu/abs/1991ApJS...77...67S}
}

@ARTICLE{Raiteri2015,
       author = {{Raiteri}, C.~M. and {Stamerra}, A. and {Villata}, M. and {Larionov}, V.~M. and {Acosta-Pulido}, J.~A. and {Ar{\'e}valo}, M.~J. and {Arkharov}, A.~A. and {Bachev}, R. and {Ben{\'\i}tez}, E. and {Bozhilov}, V.~V. and {Borman}, G.~A. and {Buemi}, C.~S. and {Calcidese}, P. and {Carnerero}, M.~I. and {Carosati}, D. and {Chigladze}, R.~A. and {Damljanovic}, G. and {Di Paola}, A. and {Doroshenko}, V.~T. and {Efimova}, N.~V. and {Ehgamberdiev}, Sh. A. and {Giroletti}, M. and {Gonz{\'a}lez-Morales}, P.~A. and {Grinon-Marin}, A.~B. and {Grishina}, T.~S. and {Hiriart}, D. and {Ibryamov}, S. and {Klimanov}, S.~A. and {Kopatskaya}, E.~N. and {Kurtanidze}, O.~M. and {Kurtanidze}, S.~O. and {Kurtenkov}, A.~A. and {Larionova}, L.~V. and {Larionova}, E.~G. and {L{\'a}zaro}, C. and {L{\"a}hteenm{\"a}ki}, A. and {Leto}, P. and {Markovic}, G. and {Mirzaqulov}, D.~O. and {Mokrushina}, A.~A. and {Morozova}, D.~A. and {M{\'u}jica}, R. and {Nazarov}, S.~V. and {Nikolashvili}, M.~G. and {Ohlert}, J.~M. and {Ovcharov}, E.~P. and {Paiano}, S. and {Pastor Yabar}, A. and {Prandini}, E. and {Ramakrishnan}, V. and {Sadun}, A.~C. and {Semkov}, E. and {Sigua}, L.~A. and {Strigachev}, A. and {Tammi}, J. and {Tornikoski}, M. and {Trigilio}, C. and {Troitskaya}, Yu. V. and {Troitsky}, I.~S. and {Umana}, G. and {Velasco}, S. and {Vince}, O.},
        title = "{The WEBT campaign on the BL Lac object PG 1553+113 in 2013. An analysis of the enigmatic synchrotron emission}",
      journal = {\mnras},
         year = 2015,
        month = nov,
       volume = {454},
       number = {1},
        pages = {353-367},
          doi = {10.1093/mnras/stv1884},
archivePrefix = {arXiv},
       eprint = {1509.02706},
 primaryClass = {astro-ph.HE},
       adsurl = {https://ui.adsabs.harvard.edu/abs/2015MNRAS.454..353R}
}

@ARTICLE{Bessell1998,
       author = {{Bessell}, M.~S. and {Castelli}, F. and {Plez}, B.},
        title = "{Model atmospheres broad-band colors, bolometric corrections and temperature calibrations for O - M stars}",
      journal = {\aap},
         year = 1998,
        month = may,
       volume = {333},
        pages = {231-250},
       adsurl = {https://ui.adsabs.harvard.edu/abs/1998A&A...333..231B}
}

@ARTICLE{Howell1988,
       author = {{Howell}, S.~B. and {Mitchell}, K.~J. and {Warnock}, III, A.},
        title = "{Statistical error analysis in CCD time-resolved photometry with applications to variable stars and quasars.}",
      journal = {\aj},
         year = 1988,
        month = jan,
       volume = {95},
        pages = {247-256},
          doi = {10.1086/114634},
       adsurl = {https://ui.adsabs.harvard.edu/abs/1988AJ.....95..247H}
}

@ARTICLE{Goyal2013,
       author = {{Goyal}, Arti and {Gopal-Krishna}, Paul J., Wiita and {Stalin}, C.~S. and {Sagar}, Ram},
        title = "{Improved characterization of intranight optical variability of prominent AGN classes}",
      journal = {\mnras},
         year = 2013,
        month = oct,
       volume = {435},
       number = {2},
        pages = {1300-1312},
          doi = {10.1093/mnras/stt1373},
archivePrefix = {arXiv},
       eprint = {1307.5831},
 primaryClass = {astro-ph.HE},
       adsurl = {https://ui.adsabs.harvard.edu/abs/2013MNRAS.435.1300G}
}

@ARTICLE{Villforth2010,
       author = {{Villforth}, Carolin and {Koekemoer}, Anton M. and {Grogin}, Norman A.},
        title = "{A New Extensive Catalog of Optically Variable Active Galactic Nuclei in the GOODS Fields and a New Statistical Approach to Variability Selection}",
      journal = {\apj},
         year = 2010,
        month = nov,
       volume = {723},
       number = {1},
        pages = {737-754},
          doi = {10.1088/0004-637X/723/1/737},
archivePrefix = {arXiv},
       eprint = {1008.3384},
 primaryClass = {astro-ph.CO},
       adsurl = {https://ui.adsabs.harvard.edu/abs/2010ApJ...723..737V}
}

@ARTICLE{Heidt1996,
       author = {{Heidt}, J. and {Wagner}, S.~J.},
        title = "{Statistics of optical intraday variability in a complete sample of radio-selected BL Lacertae objects.}",
      journal = {\aap},
         year = 1996,
        month = jan,
       volume = {305},
        pages = {42},
          doi = {10.48550/arXiv.astro-ph/9506032},
archivePrefix = {arXiv},
       eprint = {astro-ph/9506032},
 primaryClass = {astro-ph},
       adsurl = {https://ui.adsabs.harvard.edu/abs/1996A&A...305...42H}
}

@ARTICLE{Chang2024,
       author = {{Chang}, X. and {Xiong}, D.~R. and {Yi}, T.~F. and {Liu}, C.~X. and {Bhatta}, G. and {Xu}, J.~R. and {Gong}, Y.~L.},
        title = "{Optical intraday variability analysis for the BL Lacertae object 1ES 1426+42.8}",
      journal = {\mnras},
         year = 2024,
        month = sep,
       volume = {533},
       number = {1},
        pages = {120-130},
          doi = {10.1093/mnras/stae1839},
archivePrefix = {arXiv},
       eprint = {2409.06983},
 primaryClass = {astro-ph.GA},
       adsurl = {https://ui.adsabs.harvard.edu/abs/2024MNRAS.533..120C}
}

@ARTICLE{Agarwal2021,
       author = {{Agarwal}, A. and {Mihov}, B. and {Andruchow}, I. and {Cellone}, S.~A. and {Anupama}, G.~C. and {Agrawal}, V. and {Zola}, S. and {Slavcheva-Mihova}, L. and {{\"O}zd{\"o}nmez}, A. and {Ege}, Erg{\"u}n and {Raj}, A. and {Mammana}, L. and {Zibecchi}, L. and {Fern{\'a}ndez-Laj{\'u}s}, E.},
        title = "{Multi-band behaviour of the TeV blazar PG 1553+113 in optical range on diverse timescales. Flux and spectral variations}",
      journal = {\aap},
         year = 2021,
        month = jan,
       volume = {645},
          eid = {A137},
        pages = {A137},
          doi = {10.1051/0004-6361/202039301},
archivePrefix = {arXiv},
       eprint = {2011.04074},
 primaryClass = {astro-ph.HE},
       adsurl = {https://ui.adsabs.harvard.edu/abs/2021A&A...645A.137A}
}

@ARTICLE{Pandey2019,
       author = {{Pandey}, Ashwani and {Gupta}, Alok C. and {Wiita}, Paul J. and {Tiwari}, S.~N.},
        title = "{Optical Flux and Spectral Variability of the TeV Blazar PG 1553+113}",
      journal = {\apj},
         year = 2019,
        month = feb,
       volume = {871},
       number = {2},
          eid = {192},
        pages = {192},
          doi = {10.3847/1538-4357/aaf974},
archivePrefix = {arXiv},
       eprint = {1901.06696},
 primaryClass = {astro-ph.HE},
       adsurl = {https://ui.adsabs.harvard.edu/abs/2019ApJ...871..192P}
}

@ARTICLE{Valtaoja1999,
       author = {{Valtaoja}, E. and {L{\"a}hteenm{\"a}ki}, A. and {Ter{\"a}sranta}, H. and {Lainela}, M.},
        title = "{Total Flux Density Variations in Extragalactic Radio Sources. I. Decomposition of Variations into Exponential Flares}",
      journal = {\apjs},
         year = 1999,
        month = jan,
       volume = {120},
       number = {1},
        pages = {95-99},
          doi = {10.1086/313170},
       adsurl = {https://ui.adsabs.harvard.edu/abs/1999ApJS..120...95V}
}

@ARTICLE{Foschini2011,
       author = {{Foschini}, L. and {Ghisellini}, G. and {Tavecchio}, F. and {Bonnoli}, G. and {Stamerra}, A.},
        title = "{Search for the shortest variability at gamma rays in flat-spectrum radio quasars}",
      journal = {\aap},
         year = 2011,
        month = jun,
       volume = {530},
          eid = {A77},
        pages = {A77},
          doi = {10.1051/0004-6361/201117064},
archivePrefix = {arXiv},
       eprint = {1101.1085},
 primaryClass = {astro-ph.HE},
       adsurl = {https://ui.adsabs.harvard.edu/abs/2011A&A...530A..77F}
      }

@ARTICLE{Chang2023,
       author = {{Chang}, X. and {Yi}, T.~F. and {Xiong}, D.~R. and {Liu}, C.~X. and {Yang}, X. and {Li}, H.~Z. and {Gong}, Y.~L. and {Na}, W.~W. and {Li}, Y. and {Chen}, Z.~H. and {Chen}, J.~P. and {Mao}, L.~S.},
        title = "{Multicolour optical variability monitoring of blazars with high time resolution}",
      journal = {\mnras},
         year = 2023,
        month = apr,
       volume = {520},
       number = {3},
        pages = {4118-4133},
          doi = {10.1093/mnras/stad409},
archivePrefix = {arXiv},
       eprint = {2302.01420},
 primaryClass = {astro-ph.HE},
       adsurl = {https://ui.adsabs.harvard.edu/abs/2023MNRAS.520.4118C}
}

@ARTICLE{Gaur2012,
       author = {{Gaur}, Haritma and {Gupta}, Alok C. and {Strigachev}, A. and {Bachev}, R. and {Semkov}, E. and {Wiita}, Paul J. and {Peneva}, S. and {Boeva}, S. and {Kacharov}, N. and {Mihov}, B. and {Ovcharov}, E.},
        title = "{Quasi-simultaneous two-band optical variability of the blazars 1ES 1959+650 and 1ES 2344+514}",
      journal = {\mnras},
         year = 2012,
        month = mar,
       volume = {420},
       number = {4},
        pages = {3147-3162},
          doi = {10.1111/j.1365-2966.2011.20243.x},
archivePrefix = {arXiv},
       eprint = {1112.3125},
 primaryClass = {astro-ph.CO},
       adsurl = {https://ui.adsabs.harvard.edu/abs/2012MNRAS.420.3147G}
}

@ARTICLE{Yuan2015,
       author = {{Yuan}, Y.~H. and {Fan}, J.~H. and {Pan}, H.~J.},
        title = "{Optical Photometry of the BL Lac Object 1ES 1959+650}",
      journal = {\aj},
         year = 2015,
        month = sep,
       volume = {150},
       number = {3},
          eid = {67},
        pages = {67},
          doi = {10.1088/0004-6256/150/3/67},
       adsurl = {https://ui.adsabs.harvard.edu/abs/2015AJ....150...67Y}
}

@ARTICLE{Paliya2017,
       author = {{Paliya}, Vaidehi S. and {Stalin}, C.~S. and {Ajello}, M. and {Kaur}, A.},
        title = "{Intra-night Optical Variability Monitoring of Fermi Blazars: First Results from 1.3 m J. C. Bhattacharya Telescope}",
      journal = {\apj},
         year = 2017,
        month = jul,
       volume = {844},
       number = {1},
          eid = {32},
        pages = {32},
          doi = {10.3847/1538-4357/aa77f5},
archivePrefix = {arXiv},
       eprint = {1812.10614},
 primaryClass = {astro-ph.HE},
       adsurl = {https://ui.adsabs.harvard.edu/abs/2017ApJ...844...32P}
}

@INPROCEEDINGS{Currie2014,
       author = {{Currie}, M.~J. and {Berry}, D.~S. and {Jenness}, T. and {Gibb}, A.~G. and {Bell}, G.~S. and {Draper}, P.~W.},
        title = "{Starlink Software in 2013}",
    booktitle = {Astronomical Data Analysis Software and Systems XXIII},
         year = 2014,
       editor = {{Manset}, N. and {Forshay}, P.},
       series = {Astronomical Society of the Pacific Conference Series},
       volume = {485},
        month = may,
        pages = {391},
       adsurl = {https://ui.adsabs.harvard.edu/abs/2014ASPC..485..391C}
}

@INPROCEEDINGS{Berry2022,
       author = {{Berry}, David and {Graves}, Sarah and {Bell}, Graham S. and {Currie}, Malcolm and {Draper}, Peter and {Jenness}, Tim},
        title = "{Starlink - The Original and Best}",
    booktitle = {Astronomical Data Analysis Software and Systems XXX},
         year = 2022,
       editor = {{Ruiz}, Jose Enrique and {Pierfedereci}, Francesco and {Teuben}, Peter},
       series = {Astronomical Society of the Pacific Conference Series},
       volume = {532},
        month = jul,
        pages = {559},
       adsurl = {https://ui.adsabs.harvard.edu/abs/2022ASPC..532..559B}
}

@INPROCEEDINGS{Bell2024,
       author = {{Bell}, G.~S. and {Berry}, D.~S. and {Graves}, S.~F. and {Currie}, M.~J. and {Draper}, P.~W.},
        title = "{Starlink: the 2021A release}",
    booktitle = {Astronomical Data Analysis Software and Systems XXXI},
         year = 2024,
       editor = {{Hugo}, B.~V. and {Van Rooyen}, R. and {Smirnov}, O.~M.},
       series = {Astronomical Society of the Pacific Conference Series},
       volume = {535},
        month = may,
        pages = {455},
       adsurl = {https://ui.adsabs.harvard.edu/abs/2024ASPC..535..455B}
}

@ARTICLE{Abdo2010sed,
       author = {{Abdo}, A.~A. and {Ackermann}, M. and {Agudo}, I. and {Ajello}, M. and {Aller}, H.~D. and {Aller}, M.~F. and {Angelakis}, E. and {Arkharov}, A.~A. and {Axelsson}, M. and {Bach}, U. and {Baldini}, L. and {Ballet}, J. and {Barbiellini}, G. and {Bastieri}, D. and {Baughman}, B.~M. and {Bechtol}, K. and {Bellazzini}, R. and {Benitez}, E. and {Berdyugin}, A. and {Berenji}, B. and {Blandford}, R.~D. and {Bloom}, E.~D. and {Boettcher}, M. and {Bonamente}, E. and {Borgland}, A.~W. and {Bregeon}, J. and {Brez}, A. and {Brigida}, M. and {Bruel}, P. and {Burnett}, T.~H. and {Burrows}, D. and {Buson}, S. and {Caliandro}, G.~A. and {Calzoletti}, L. and {Cameron}, R.~A. and {Capalbi}, M. and {Caraveo}, P.~A. and {Carosati}, D. and {Casandjian}, J.~M. and {Cavazzuti}, E. and {Cecchi}, C. and {{\c{C}}elik}, {\"O}. and {Charles}, E. and {Chaty}, S. and {Chekhtman}, A. and {Chen}, W.~P. and {Chiang}, J. and {Chincarini}, G. and {Ciprini}, S. and {Claus}, R. and {Cohen-Tanugi}, J. and {Colafrancesco}, S. and {Cominsky}, L.~R. and {Conrad}, J. and {Costamante}, L. and {Cutini}, S. and {D'ammando}, F. and {Deitrick}, R. and {D'Elia}, V. and {Dermer}, C.~D. and {de Angelis}, A. and {de Palma}, F. and {Digel}, S.~W. and {Donnarumma}, I. and {Silva}, E. do Couto e. and {Drell}, P.~S. and {Dubois}, R. and {Dultzin}, D. and {Dumora}, D. and {Falcone}, A. and {Farnier}, C. and {Favuzzi}, C. and {Fegan}, S.~J. and {Focke}, W.~B. and {Forn{\'e}}, E. and {Fortin}, P. and {Frailis}, M. and {Fuhrmann}, L. and {Fukazawa}, Y. and {Funk}, S. and {Fusco}, P. and {G{\'o}mez}, J.~L. and {Gargano}, F. and {Gasparrini}, D. and {Gehrels}, N. and {Germani}, S. and {Giebels}, B. and {Giglietto}, N. and {Giommi}, P. and {Giordano}, F. and {Giuliani}, A. and {Glanzman}, T. and {Godfrey}, G. and {Grenier}, I.~A. and {Gronwall}, C. and {Grove}, J.~E. and {Guillemot}, L. and {Guiriec}, S. and {Gurwell}, M.~A. and {Hadasch}, D. and {Hanabata}, Y. and {Harding}, A.~K. and {Hayashida}, M. and {Hays}, E. and {Healey}, S.~E. and {Heidt}, J. and {Hiriart}, D. and {Horan}, D. and {Hoversten}, E.~A. and {Hughes}, R.~E. and {Itoh}, R. and {Jackson}, M.~S. and {J{\'o}hannesson}, G. and {Johnson}, A.~S. and {Johnson}, W.~N. and {Jorstad}, S.~G. and {Kadler}, M. and {Kamae}, T. and {Katagiri}, H. and {Kataoka}, J. and {Kawai}, N. and {Kennea}, J. and {Kerr}, M. and {Kimeridze}, G. and {Kn{\"o}dlseder}, J. and {Kocian}, M.~L. and {Kopatskaya}, E.~N. and {Koptelova}, E. and {Konstantinova}, T.~S. and {Kovalev}, Y.~Y. and {Kovalev}, Yu. A. and {Kurtanidze}, O.~M. and {Kuss}, M. and {Lande}, J. and {Larionov}, V.~M. and {Latronico}, L. and {Leto}, P. and {Lindfors}, E. and {Longo}, F. and {Loparco}, F. and {Lott}, B. and {Lovellette}, M.~N. and {Lubrano}, P. and {Madejski}, G.~M. and {Makeev}, A. and {Marchegiani}, P. and {Marscher}, A.~P. and {Marshall}, F. and {Max-Moerbeck}, W. and {Mazziotta}, M.~N. and {McConville}, W. and {McEnery}, J.~E. and {Meurer}, C. and {Michelson}, P.~F. and {Mitthumsiri}, W. and {Mizuno}, T. and {Moiseev}, A.~A. and {Monte}, C. and {Monzani}, M.~E. and {Morselli}, A. and {Moskalenko}, I.~V. and {Murgia}, S. and {Nestoras}, I. and {Nilsson}, K. and {Nizhelsky}, N.~A. and {Nolan}, P.~L. and {Norris}, J.~P. and {Nuss}, E. and {Ohsugi}, T. and {Ojha}, R. and {Omodei}, N. and {Orlando}, E. and {Ormes}, J.~F. and {Osborne}, J. and {Ozaki}, M. and {Pacciani}, L. and {Padovani}, P. and {Pagani}, C. and {Page}, K. and {Paneque}, D. and {Panetta}, J.~H. and {Parent}, D. and {Pasanen}, M. and {Pavlidou}, V. and {Pelassa}, V. and {Pepe}, M. and {Perri}, M. and {Pesce-Rollins}, M. and {Piranomonte}, S. and {Piron}, F. and {Pittori}, C. and {Porter}, T.~A. and {Puccetti}, S. and {Rahoui}, F. and {Rain{\`o}}, S. and {Raiteri}, C. and {Rando}, R. and {Razzano}, M. and {Reimer}, A. and {Reimer}, O.},
        title = "{The Spectral Energy Distribution of Fermi Bright Blazars}",
      journal = {\apj},
         year = 2010,
        month = jun,
       volume = {716},
       number = {1},
        pages = {30-70},
          doi = {10.1088/0004-637X/716/1/30},
archivePrefix = {arXiv},
       eprint = {0912.2040},
 primaryClass = {astro-ph.CO},
       adsurl = {https://ui.adsabs.harvard.edu/abs/2010ApJ...716...30A}
}

\end{document}